\documentstyle[12pt,preprint]{aastex}
\begin{document}
\shorttitle{$B$,$V$ Photometry of NE arm SMC Variable Stars}
\shortauthors{Sharpee et al.}
\title{$B$,$V$ Photometry of Variable Stars in the Northeast Arm of the Small
Magellanic Cloud}
\author{Brian Sharpee, Michele Stark\altaffilmark{1}, Barton Pritzl
\altaffilmark{2},
\and 
Horace Smith\altaffilmark{3}}  
\affil{Department of Physics and Astronomy, Michigan State University, 
East Lansing, MI 48824}
\email{smith@pa.msu.edu}

\author{Nancy Silbermann\altaffilmark{3}}
\affil{SIRTF Science Center, California Institute of Technology, \\
MS 220-6, Pasadena, CA 91125}
\email{nancys@ipac.caltech.edu}

\author{Ronald Wilhelm}
\affil{Department of Physics, Texas Tech University, Lubbock, TX 79409}
\email{ron.wilhelm@ttu.edu}

\author{Alistair Walker}
\affil{Cerro Tololo Interamerican Observatory, National Optical
Astronomy Observatories, \\ P.O. Box 26732, Tucson, AZ 85726}
\email{awalker@noao.edu}

\altaffiltext{1}{Current address: 525 Davey Lab, Department of Astronomy and
Astrophysics, Pennsylvania
State University, University Park, PA 16802}

\altaffiltext{2}{Current address: National Optical Astronomy Observatories, 
P.O. Box 26732, Tucson, AZ 85726 \verb+pritzl@noao.edu+} 

\altaffiltext{3}{Visiting Astronomer, Cerro Tololo Inter-American
Observatory, National Optical Astronomy Observatories, which is
operated by AURA, Inc., under cooperative agreement with the
National Science Foundation}

\begin{abstract}
$B$ and $V$ photometry has been obtained for variable stars in the
northeast arm of the Small Magellanic Cloud (SMC).\@ Periods and
lightcurves have been determined for 237 periodic variables, including
201 Cepheids, 68 of which are newly discovered.  Fundamental mode
Cepheids and first overtone mode Cepheids are generally well separated
in the $B,V$ color-magnitude diagram, with the latter having bluer
mean colors than the former.  The Cepheid period-color relationship
for this outlying SMC field is indistinguishable from that seen in
more centrally located SMC fields, and is bluer than theoretical
predictions.  The red edge to the populated portion of the instability
strip shifts to bluer colors for fainter Cepheids. There is support
from our sample for a previously reported steepening in the slope of
the period-luminosity relation for fundamental mode Cepheids near a
period of 2 days.  The Cepheids of the northeast arm may be closer to
us than are those of the main body of the SMC, but the difference is
smaller than or equal to about 4 kpc, comparable to the tidal radius
of the SMC.

\end{abstract}

\keywords{Stars:Variables:Cepheids; Galaxies: Magellanic Clouds}

\section{Introduction}

Leavitt \citep{P12} discovered the period-luminosity relation using
observations of Cepheid variable stars in the Small Magellanic Cloud
(SMC).  Because the young population of the SMC is now known to be
metal-poor relative to that of the solar neighborhood and to that of
the Large Magellanic Cloud, SMC Cepheids remain of particular interest
in the elucidation of the effects of metallicity upon the properties
of Cepheid variable stars.  Moreover, it has been proposed that the
depth of the SMC along the line of sight is not insignificant compared
to its mean distance (Westerlund 1997 and references therein).
Cepheids provide a means of measuring the distances to the young
populations of different parts of the SMC.

\citet{S92} used photographic photometry in the $B$ band to study the
variable star population in an outlying field in the northeast arm of
the SMC centered on RA = $1^{h}03^{m}$ and DEC = $-71\arcdeg
23\arcmin$ (J2000). The brighter variable stars within this field have
been studied by \citet{A60} and \citet{PGG66}, who concentrated
primarily on the variable stars discovered previously at Harvard.
Extending the search for variable stars to fainter limits, \citet{S92}
identified 136 new variables in the area and obtained photographic $B$
light curves for 150 new and old variables.  Their period-luminosity
diagram showed the presence of both fundamental mode and first
overtone mode Cepheids, but, because they observed only in the $B$
band, they could not address questions regarding the colors of the
variable stars they observed.

We have obtained $B$ and $V$ CCD photometry of variable stars in this
field, including 84 newly discovered variables, 63 variables first
observed in \citet{S92}, 2 variables present in the \citet{C01}
catalog of globular cluster variable stars, as well as 88 Harvard
variables.  Our purpose in carrying out this survey was to address
several issues.  First, the present CCD survey more readily detects
low amplitude Cepheids than did the earlier photographic study,
providing a more complete picture of the variable star population of
the northeast arm.  Second, the new data allow us to discover where
the fundamental mode and first overtone mode Cepheids fall within the
color-magnitude diagram (CMD).  In particular, is there overlap beyond
that attributable to observational error in the colors of Cepheids
having different pulsation modes, or do fundamental mode and first
overtone mode Cepheids occupy distinct and separate regions in the
CMD?  A substantial overlapping of the fundamental mode and first
overtone mode Cepheid domains is visible in the $I_0, (V-I)_0$ CMD
based upon data taken in the {\sc ogle ii} survey (Figure 4 of Udalski
et al.\ 1999b).  However, that survey emphasized the crowded central
regions of the SMC, and problems with crowding, blended images, and
possibly differential reddening may have contributed to the overlap.
If that overlapping is real, and not attributable to observational
error, then there is no unique dividing line between fundamental mode
and first overtone mode SMC Cepheids in the HR diagram. The existence
of a hysteresis zone, in which pulsation mode depends upon direction
of evolution in the HR diagram, would be one possible way by which
such an overlap zone might be produced.  Such a hysteresis zone has
been proposed for RR Lyrae variables (e.g., Stellingwerf 1975).

In addition, prior studies have found a break in the slope of the
period-luminosity ($P-L$) relation for fundamental mode SMC Cepheids
near a period of 2 days (see \S~\ref{PL}).  This break has not yet
been satisfactorily explained, and it is of interest to determine
whether the location of the short period Cepheids in the CMD can shed
light upon this problem.  We also want to compare the
period-color relations for Cepheids in the northeast arm
with those which have been derived for Cepheids in more central
regions of the SMC and with theoretical period-color relations.

Finally, we wish to compare the $P-L$ relations for the
northeast arm Cepheids with those derived in the course of
microlensing surveys of more central SMC fields.  If the slopes
of the $P-L$ relations derived for Cepheids in the northeast arm
are significantly different from those of Cepheids in the main
body of the SMC, that would indicate that the Cepheid population
of the SMC is heterogeneous, with properties depending upon location.
Alternatively, a difference only in the zero-points of the
$P-L$ relations might indicate that the Cepheids of the northeast
arm are nearer or more distant than those in other parts of the
SMC.
 
\section{Observations and Reductions} 
 
 Four fields in the northeast arm of the SMC were imaged with the
 Thomson 1024 $\times$ 1024 CCD camera on the CTIO 60-cm Curtis
 Schmidt telescope.  Each field measures 0.5 degrees on a side,
 together defining a square 1$\arcdeg$ $\times$ 1$\arcdeg$ area of the
 SMC centered approximately on RA = $1^{h}03^{m}$ and DEC =
 $-71\arcdeg 23\arcmin$ (J2000), including most of the field studied
 photographically by \citet{S92}.  There are small areas of overlap
 between our adjoining fields.  Our pixel size was approximately
 $2\arcsec$ square.  For comparison, the pixel size for the detector
 utilized by the {\sc ogle} group in their microlensing survey ({\sc
 ogle ii}), to which comparisons are made in this paper, was
 approximately $0.4\arcsec$ square, although they were limited to a
 seeing resolution of about $1\arcsec$ in the visible bands due to
 non-uniformities in telescope tracking complicating the
 driftscan-mode acquisition of their detector \citep{U97}.
 
 The four fields were observed with Johnson $B$ and $V$ filters during
 five runs in 1992, 1993, and 1994.  A journal of observations for
 these observing runs is provided in Table~\ref{tbl1}. During the
 first runs in September and October, 1992, the usual observing
 procedure was to obtain three consecutive $V$ and four consecutive
 $B$ exposures of each field, each 300~s long.  These were later
 combined to constitute one 900~s $V$ and one 1200~s $B$ observation.
 During runs in September and November, 1993, and October, 1994, the
 exposure times in $B$ were increased to 420 seconds, so that the
 total exposure from the combination of four consecutive $B$ images
 was 1680~s. Because of the poorer blue sensitivity of the Thomson
 chip, the $B$ images do not reach as faint as the $V$ despite the
 longer exposure times.
 
 Images were flattened with sky flats, and conventionally reduced with
 {\sc iraf} \footnote{IRAF is distributed by the National Optical
 Astronomy Observatories, which are operated by the Association of
 Universities for Research in Astronomy, Inc., under cooperative
 agreement with the National Science Foundation.}.  Aperture
 photometry was obtained with the standalone version of {\sc daophot
 ii} \citep{S87}.  Because the CCD pixels are 2$\arcsec$ square,
 confusion of images can be a significant source of error in the more
 crowded portions of the fields.
 
 The photometry for the individual frames was first reduced to an
 adopted instrumental magnitude system.  Possible variable stars were
 then identified from the scatter about the mean value in the
 instrumental $b$ and $v$ magnitudes.  Examination of the images of
 suspected variables eliminated those stars for which the scatter
 could be attributed to crowding or other causes not related to true
 variability. Typically, each variable was observed about 22 times in
 each filter, although a few variables in the overlap zones between
 fields were observed more frequently.  Quantitatively estimating the
 completeness of this survey is difficult, given stars are lost due to
 magnitude, amplitude, and period considerations as well as confusion
 in crowded images.  Nevertheless, Cepheid variable stars with periods
 under 20 days, amplitudes greater than about 0.2 mag, apparent
 magnitudes brighter than $\langle{V}\rangle$ = 18.5, and which were
 not subject to severe crowding, should have been detected in this
 search.  The completeness of variable star identification declines
 sharply at magnitudes fainter than 18.5, even for uncrowded stars.
 The identification of SMC RR Lyrae variables, at
 $\langle{V}\rangle\sim19.5$, is very incomplete, and only 5-10\% of
 the SMC RR Lyrae stars within the field are likely to have been
 detected.  Completeness also declines in the most crowded regions of
 our fields, particularly affecting the most southerly areas of the
 survey. We still would expect to identify most SMC Cepheids of
 $\langle{V}\rangle \leq 17.5$, even in crowded fields, although the
 measured light curves might not be of high quality.
 
 A period search was conducted for the suspected variables using the
 phase dispersion minimization program, as implemented in {\sc iraf}.
 The spacing of our observations is well-suited for determining
 periods for stars which have periods shorter than about 5 days, but
 is not well suited for longer periods.  Determinations of periods
 longer than 20 days are not possible with any reliability.
 \citet{HW77}, \citet{S92}, and \citet{C01} were consulted to identify
 known variable stars within our fields.  Although the periods for
 most of these previously discovered variables are already known, a
 new period search was nonetheless carried out for each of them.

Newly discovered variables are identified in
Figures~\ref{fig1a}-\ref{fig1d}.  Positions for the new variables were
calculated on the system of the Digitized Sky Survey, using routines
in {\sc iraf}.  The resultant positions, good to about 2$\arcsec$, are
listed in Table~\ref{tbl2}.  The error was estimated from comparisons
of the positions we derived with those listed in the MACS catalog
\citep{T96} for stars in common.  The numbering scheme for the new
variables is prefixed with ``SSP'' and numbers 84 stars.
 
Local standards were set up in a single $13.6 \arcmin \times 13.6
\arcmin$ field, centered on RA = 01$^{h}$04$^{m}$49$^{s}$, Dec =
--71$\arcdeg23\arcmin00\arcsec$ (J2000), with the CTIO 0.9-m telescope
and CCD imaging system on 8 November 1999.  CCD SITe 2048 \#3 was used
together with the standard 0.9-m $UBVRI$ filter set.  Observations of
the standard field were made in $U,B,V,I$ filters, of which only $B$ and
$V$ are relevant for the calibration purposes required here.  Exposure
times were 2 $\times$ 100s, 500s ($V$) and 2 $\times$ 200s, 1000s
($B$).  Twilight sky flat field exposures successfully reduced
systematics to well below one percent.

 The night was photometric and observations were made of 105
\citet{L92} standards in 10 fields at a variety of airmasses ranging
from 1.07--1.85.  Of these stars a few, with large previously measured
residuals \citep{L92} or which fell into CCD defects, were rejected.
However, 86 were retained in the final color equation fits, which
confirmed the high quality of the night, with zeropoints constrained
to 0.0015 -- 0.002 mag.

A set of 21 local standards, selected as being isolated stars covering
a wide magnitude and color range ($V:12-17, B-V:-0.2-+1.5$) were
measured using aperture photometry in exactly the same way as the
primary standard stars, in airmasses between 1.33 and 1.35.  Stars
brighter than magnitude 13.5 were measured on the short exposures
only; the remaining stars all have three measurements.  Agreement
between measurements for any given star is typically 0.01
mag. Photometry and positions for the local standards are listed in
Table~\ref{tbl3}. Results are listed individually for the two shorter
and one longer exposures. Values of right ascension and declination
were found in the same fashion as for the new variable stars.  The
local standard stars are identified in Figure~\ref{fig2}.

 Instrumental $b$ and $v$ magnitudes were transformed to the Johnson
 $B$ and $V$ systems using the local standard stars.  Transformation
 equations had the form:
 \begin{eqnarray}
 b = B + C_B - 0.05(B-V)\,, \\
 v = V + C_V - 0.01(B-V)\,.
 \end{eqnarray}
 
 Two existing photometric sequences were used to check the accuracy of
 the transformation. Our two more northern fields include stars near
 \objectname{NGC 362} which were observed photoelectrically and
 photographically in $B$ and $V$ by \citet{H82}.  Secondly, one of our
 fields includes the SMC cluster \objectname{NGC 411}, for which
 \citet{DCM86} obtained $B,R$ photometry. Mean differences between our
 $B$ and $V$ values and those in \citet{H82} and \citet{DCM86} are
 summarized in Table~\ref{tbl4}.  The number of stars compared and the
 mean differences in the magnitudes, in the sense their values minus
 ours, are listed in this table.  The different photometric systems
 appear to agree to within about $\pm$~0.02 mag.  The photometry for
 individual variable stars is listed in Table~\ref{tbl5}.  The full
 table will be given in the electronic version of this paper.  The
 errors listed in column 4 of this table are the formal standard
 errors returned by the {\sc daophot ii} code, and may in some cases
 underestimate the actual uncertainty.
 
\section{Lightcurves, Periods, and Mean Colors}

Data for the periodic variables observed in this survey are summarized
in Table~\ref{tbl6}.  The variable is identified in column (1) by its
Harvard Variable number (HV), its designation in the \citet{C01}
catalog (NGC362V), its number in \citet{S92} ([SSB92]), or by its
number in Table~\ref{tbl2} (SSP).  The variable type is listed in
column (2), where ``F'' indicates a Cepheid pulsating in the
fundamental mode, ``O'' indicates a Cepheid pulsating in the first
overtone mode, ``RRab'' indicates an RR Lyrae star in the fundamental
mode, ``RRc'' indicates an RR Lyrae star in the first overtone mode,
and ``E'' indicates an eclipsing variable star.  Our derived period,
typically good to a few parts in 10$^5$, is given in column (3).  We
list luminosity-weighted mean $\langle{B}\rangle$ and
$\langle{V}\rangle$ magnitudes in columns (4) and (5).
Luminosity-weighted $\langle{B}\rangle - \langle{V}\rangle$ and
magnitude-weighted mean $(\bv)_{mag}$ colors are listed in columns (6)
and (7).  In column (8) a ``1'' signifies a variable with a relatively
complete and well defined lightcurve, whereas a ``2'' indicates that
the lightcurve is noisy or has substantial gaps.  In general, only
stars with quality ``1'' lightcurves were selected for more detailed
analysis.  For a few variables, a good lightcurve was obtained in $B$
or $V$, but not in both bandpasses.  These are indicated by a
``1($V$)'' or ``1($B$)'' in column (8).  A cubic spline was fit to
each lightcurve and used to calculate all mean magnitudes and colors.

The decision to assign a Cepheid to the first overtone or fundamental
mode category was made by inspection of the lightcurves.  Previous
studies have indicated that, at a given period, the lightcurves of
first overtone pulsators are characterized by smaller amplitudes and
less asymmetry than those of fundamental mode pulsators (e.g.\ Smith
et al.\ 1992).  Classifications of pulsation mode based on lightcurves
appear to be fully consistent with the locations of the Cepheids in
the period-luminosity domain.  The number of data points for most of
the Cepheids is not large enough to effectively apply Fourier
decomposition methods for determining pulsation mode.  It is also
possible that some of the Cepheids, particularly among the quality
``2'' group with periods less than a day, may be double-mode
pulsators.  Double-mode Cepheids have been discovered in both the LMC
\citep{A99,S00} and SMC \citep{Be97}.  However, we do not have a
sufficiently large number of observations to effectively test for
double-mode behavior among these stars.

$B$ and $V$ lightcurves for periodic variables are shown in
Figure~\ref{fig3}, arranged in order of descending period.  Phases
have been calculated from the formula:

\begin{eqnarray}
\phi = (JD_{\sun}/P)\,,
\end{eqnarray}

\noindent where $JD_{\sun}$ is the full heliocentric Julian Date and
$P$ is the period in days, with only the non-integer part of the
quotient retained as the phase. Pulsating variables of all types with
quality ``1'' lightcurves in both $B$ and $V$ are plotted in the
$\langle{V}\rangle - \log P$ diagram in Figure~\ref{fig4}.

\section{Reddening}

\citet{H82} adopted $E(\bv)$ = 0.04 for the globular cluster
\objectname{NGC 362}, which is located close to our northeast arm
field.  \citet{DCM86} also adopted $E(\bv)$ = 0.04 for \objectname{NGC
411}, which is within our field.  \citet{CC86} found a mean reddening
of $E(\bv)$ = 0.054 for all except the central region of the SMC.\@
\citet{GM86} obtained $E(\bv)$ = 0.09 for SMC supergiants, with no
trend with location in the SMC.\@ Based upon these results,
\citet{S92} adopted a mean reddening of $E(\bv)$ = 0.06 for the
northeast arm region.  More recently, \citet{U99b} obtained reddening
values of $E(\bv)$ = 0.079 and 0.084 for their SMC fields SC10 and
SC11, the two fields in their survey which are closest to our own.
Both SC10 and SC11 lie nearer to the central body of the SMC than does
our field.  SC10 is centered at RA = $1^{h}04^{m}51^{s}$, DEC =
$-72\arcdeg24\arcmin25\arcsec$ (J2000), and SC11 is centered at RA =
$1^{h}07^{m}45^{s}$ and DEC = $-72\arcdeg39\arcmin30\arcsec$ (J2000).
\citet{U99b} assigned $E(\bv)$ = 0.070 to their field SC1 which,
although on the sky it lies on the other side of the SMC from the
northeast arm, may be the {\sc ogle ii} field most similar to our own in
terms of stellar density.

We can also estimate the foreground reddening in the direction of the
northeast arm using the color data for Galactic RR Lyrae stars within
our field.  \citet{S66} and \citet{B92} determined that the intrinsic
colors of ab type RR Lyrae stars were a function only of metallicity,
period, and average color in the phase interval $0.5 \le \phi \le 0.8$
after maximum light.  Here we employ the relationship, found by
\citet{B92}, that the reddening within this phase interval is given by:
\begin{eqnarray}
E(\bv)& = & {\langle\bv\rangle}_{\phi(0.5-0.8)} + 0.0122
 \Delta S \nonumber \\ & & - 0.00045 (\Delta S)^{2} - 0.185 P - 0.356
 \,,
\end{eqnarray}
\noindent where $P$ is the period in days and $\Delta S$ is the
spectroscopic metallicity index for RR Lyrae stars \citep{P59}.
Unfortunately, we have no direct measurement of the metallicity index
$\Delta S$ for the foreground RR Lyrae stars.  An indirect estimate of
$\Delta S$ for these variables can be made from the period shift of
the star, $\Delta \log P$, as defined in \citet{S82b}:
\begin{eqnarray}
\Delta \log P & = & -[\log P + 0.129 A_B + 0.088]\,,
\end{eqnarray}
\noindent where $A_B$ here is the $B$ amplitude, not the $B$ band
extinction.  $\Delta \log P$ is related to [Fe/H] by the relation
\citep{S82a}:
\begin{eqnarray}
\Delta \log P & = & 0.116[\textrm{Fe/H}] + 0.173 \,.
\end{eqnarray}
If we employ the calibration of the $\Delta S$ index from \citet{B75}: 
\begin{eqnarray}
\textrm{[Fe/H]} & = & -0.23 - 0.16 \Delta S\,,
\end{eqnarray}
$\Delta S$ can then be related to observable parameters and
substituted into the \citet{B92} formula to determine reddening.
Values of $E(\bv)$ determined in this fashion for foreground RRab
stars are listed in Table~\ref{tbl7}, along with the inferred values
of [Fe/H].  Some uncertainty attaches to this approach, particularly
as it is applied to individual stars, since $\Delta \log P$ also may
depend upon the evolution of the particular star under consideration.
The mean value of $E(\bv)$ derived in this way is 0.05$\pm$0.01.
The
formal error of the mean is undoubtedly an underestimate given the
uncertainty which attaches to our determination of the metallicities.
If our derived [Fe/H] values were systematically in error by 0.4 dex,
the derived mean value of $E(\bv)$ would change by 0.03, which is a
more realistic estimated uncertainty.  Although in principle the same
approach can be used to determine reddenings for RRab stars actually
belonging to the SMC, our lightcurves for such stars are too noisy to
permit us to use them in that fashion. A foreground reddening of
$E(\bv)$ = 0.05 is greater than the value of 0.019 $\pm$ 0.004 found
by \citet{MF80} from photometry of foreground stars in the direction
of the SMC, but is similar to the 0.04 adopted by \citet{H82} for
\objectname{NGC 362}.

For the remainder of this paper we adopt a reddening value $E(\bv)$ =
0.07 for the northeast arm field, a value slightly larger than the
reddening determined for the foreground RRab stars, and close to the
values \citet{U99b} obtained for outlying parts of the SMC.\@  The
adopted value of $E(\bv)$ may be uncertain by $\pm$0.02, an
uncertainty which encompasses the most recent determinations of the
reddening in the direction of the SMC.\@  We follow \citet{U99b} in
adopting the relations:
\begin{eqnarray}
A_B & = & 4.32\,E(\bv) \,, \\
A_V & = & 3.24\,E(\bv) \,,
\end{eqnarray}
\noindent giving $A_B$ = 0.30 and $A_V$ = 0.23.

\section{Period-Frequency Relations for Cepheids}

Table~\ref{tbl8} lists the distribution of the number of fundamental
and first overtone mode Cepheids as a function of period for variables
in the present survey and in the survey of \citet{S92}, which covered
a larger 1$\arcdeg$$ \times$ 1.3$\arcdeg$ area.  In this table we
include all variables with periods, not just those with quality ``1''
lightcurves. This comparison is shown graphically in
Figure~\ref{fig5}.  Allowing for the larger field size of the Smith et
al.\ photographic study, we see that the present survey identified a
modest but significant number of additional fundamental mode Cepheids
which were not included in \citet{S92}. In contrast, the number of
first overtone mode pulsators with periods has been approximately
doubled. This can be attributed to the greater ease of identifying low
amplitude variables using the CCD images.  Somewhat surprisingly, the
greatest number of additional first overtone Cepheids occurred in the
$ 0.3 < \log P_1 < 0.4$ bin, among the brighter first overtone
Cepheids.  Figure~\ref{fig6} compares histograms for all Cepheids with
periods and for those with quality ``1'' lightcurves.  Note that first
overtone Cepheids with quality ``1'' lightcurves tend to be
under-represented at periods close to one day.  This reflects the
difficulty in obtaining complete phase coverage for variables with one
day periods.  The quality ``2'' first overtone Cepheids do cluster
around one day for the same reason.  Likewise, lack of complete phase
coverage is responsible for the deficit of quality ``1'' lightcurves
among the longer period fundamental mode Cepheids.

The maximum number of fundamental mode Cepheids was observed in the
$\log P_0 = 0.2-0.3$ bin.  This is very close to the most common
period in the histogram (Figure 6, Udalski et al.\ 1999b) over $\log P$ for
fundamental mode Cepheids in the more central parts of the SMC studied
in the {\sc ogle ii} survey.  The period-frequency distribution for
fundamental mode Cepheids in the northeast arm is similar in general
to that seen in the areas surveyed in {\sc ogle ii}.

In our study, the first overtone mode Cepheids exhibit a broad peak,
cutting off sharply at periods longer than $\log P_1 = 0.4$.
\citet{U99b} also found a relatively broad peak in their histogram
(also their Figure 6) over $\log P_1$ for first overtone Cepheids in
the SMC.\@ However, their cutoff at longer periods is not so sharp as
is shown in Table~\ref{tbl8}, and their histogram is more strongly
peaked to periods around $\log P_1 = 0.1$.  We note, however, that
with the smaller number of first overtone Cepheids in our sample, this
difference may not be significant.

\section{Color-Magnitude Diagram}

 In Figure~\ref{fig7}, we plot luminosity-weighted,
 extinction-corrected mean $V$ magnitudes, $\langle V\rangle_0$,
 against magnitude-weighted, reddening-corrected mean colors,
 $(\bv)_{mag,0}$, for those Cepheids in Table~\ref{tbl6} which have
 quality ``1'' $B$ and $V$ lightcurves.  We choose to plot
 magnitude-weighted, rather than luminosity-weighted colors here
 since, at least for RR Lyrae stars, it has been argued that
 magnitude-weighted colors are closely related to the equilibrium
 temperatures of the stars \citep{D01}.  As is evident from
 Table~\ref{tbl6}, the magnitude-weighted and luminosity-weighted
 colors are in any case similar (though with small differences
 dependent upon lightcurve shape). The $(B-V)_{mag}$ colors tend to be
 about 0.01--0.02 mag redder than $\langle{V}\rangle -
 \langle{B}\rangle$ colors for the variables we observed.  For
 galactic Cepheids, \citet{B99} predict a difference of 0.018 mag at
 $\log P$ = 1.0.
 
 As expected, the Cepheids fall within a well-defined instability
 strip.  The first overtone Cepheids are for the most part well
 divided from the Cepheids which pulsate in the fundamental mode.
 There is some overlap at the boundary between first overtone and
 fundamental mode pulsators, but the overlap is mostly small, and can
 be accounted for by typical errors of $\pm$ 0.03 mag in
 $(\bv)_{mag}$.  A few stars are, however, significantly bluer or
 redder than would be expected on the basis of their pulsation mode.
 \citet{A60} noticed that HV1898 has an unusually blue color compared
 to other Cepheids of similar lightcurve shape and period, a result
 with which we concur.  He attributed this to the presence of an
 unresolved blue companion, a conclusion which seems very plausible.
 It is quite possible that all of the color overlap between
 fundamental mode and first overtone mode Cepheids in
 Figure~\ref{fig7} can be attributed to the combination of
 observational errors and the existence of a few instances of
 unresolved companions.  There is evidence for a narrowing of the
 Cepheid instability strip, or at least the populated region of the
 instability strip, at lower luminosities, with the red edge shifting
 to bluer colors for fundamental mode Cepheids.
 
 For comparison, Figure~\ref{fig8} shows a color-magnitude diagram in
 which points for the Cepheids in Udalski et al.'s (1999b) fields SC10
 and SC11 are added to those in our field.  We note that the {\sc
 ogle} project web page states that the photometry for the SMC
 Cepheids was revised on April 1, 2000.  We have used the revised
 photometry throughout this paper \citep{U00}.
 
 The pattern in Figure~\ref{fig8} is similar to that in
 Figure~\ref{fig7}: Fundamental mode pulsators tend to have redder
 colors than first overtone pulsators of equivalent luminosity.  There
 is more scatter than in Figure~\ref{fig7} and a greater overlap in
 color between fundamental mode and first overtone mode Cepheids.
 That can plausibly be attributed to a greater proportion of crowded
 star images or incomplete lightcurves in the somewhat more central
 SC10 and SC11 fields than in the selected data used to make
 Figure~\ref{fig7}.  One possibly significant difference between the
 two color-magnitude diagrams is the existence of a number of faint
 and relatively red first overtone pulsators at the bottom of the
 color-magnitude diagram (Figure~\ref{fig8}) from the {\sc ogle ii}
 sample which are not seen in our dataset.  The {\sc ogle ii} $B$ and $V$
 lightcurves of some of these redder first overtone stars are noisy,
 but that is not the case for all of them.  These variables have
 $\langle V\rangle_0 \sim 18$ and extend to $(\bv)_{mag,0} = 0.4$.  It
 is true that some of our shorter period first overtone Cepheids were
 excluded from Figure~\ref{fig7} and Figure~\ref{fig8}, on the basis
 that their lightcurves showed too much scatter (see
 Table~\ref{tbl6}).  However, we would expect that at least a few of
 the faint, redder first overtone variables would have been
 well-observed in our dataset were such stars as common in our field
 as they seem to be in SC10 and SC11.  We also note that there remains
 evidence for a blueward shift in the colors of the reddest
 fundamental mode Cepheids as one goes to fainter luminosities.  This
 will be discussed further in \S~\ref{PL}.

\section{Period-Color Relations}
 
We have determined least squares fits to the period-color relations
defined by the fundamental mode and first overtone mode Cepheids
within our northeast arm field.  To uncorrelate the errors in slope
and zero-point, we fitted about the median period of each
distribution.  For 70 fundamental mode Cepheids with reliable colors,
meaning quality ``1'' light curves in both bands, we obtain:
\begin{eqnarray}
(\bv)_{mag,0} = 0.27(\pm 0.03) (\log P_0 - \overline{\log P_0}) + 0.42(\pm 0.01) \,,
\end{eqnarray}
\noindent where $\overline{\log P_0} = 0.3631$, producing a standard
deviation of $\pm$ 0.05 mag to the fit.  For 58 first overtone
mode Cepheids with reliable colors we find:
\begin{eqnarray}
(\bv)_{mag,0} = 0.14(\pm 0.03) (\log P_1 - \overline{\log P_1}) + 0.31(\pm 0.01) \,,
\end{eqnarray}
\noindent where $\overline{\log P_1} = 0.1918$, yielding a standard
deviation of $\pm$ 0.04 mag to the fit.  To cast these relations in
a form more easily comparable with
theoretical period-color relations we have refitted our data without the offset
by median period, obtaining:
\begin{eqnarray}
(\bv)_{mag,0} = 0.27(\pm 0.03) \log P_0 + 0.32(\pm 0.01)\,, 
\end{eqnarray}
 \noindent for fundamental mode Cepheids
 and
\begin{eqnarray}
(\bv)_{mag,0} = 0.14(\pm 0.03) \log P_1 + 0.29(\pm 0.01)\,,
\end{eqnarray}
\noindent for first overtone mode Cepheids.
For comparison we include our least squares fits to 1250
fundamental mode Cepheids and 773 first overtone mode Cepheids in
\citet{U99b}: 
\begin{eqnarray}
 (\bv)_{mag,0} = 0.27(\pm 0.01) \log P_0 + 0.32(\pm 0.01) \,, \\
 (\bv)_{mag,0} = 0.18(\pm 0.02) \log P_1 + 0.301 (\pm 0.004) \,.
\end{eqnarray}

 The period-color relations for the fundamental mode and first
 overtone Cepheids are shown in Figures~\ref{fig9}
 and~\ref{fig10}. Our period-color relations and those defined by
 stars in the more centrally located {\sc ogle ii} fields are in
 excellent agreement for the period ranges covered by our data.  The
 agreement is not so good with the theoretical period-color relation
 for fundamental mode SMC Cepheids (Z = 0.004) taken from \citet{A99},
 which is significantly redder than the observed relations.
The slope
 of the period-color relation for fundamental mode Cepheids of Z =
 0.004 from \citet{C00} is closer to that which is observed, but the
 zero-point is slightly too red.  A reduction of 0.04 in the adopted
 value of E($B-V$) would bring the observed period-color relationship
 into good agreement with that of \citet{C00}. A change that large is
 probably unlikely, but an error of 0.02 in the adopted reddening
 value can certainly not be excluded.  The theoretical period-color
 relationships of \citet{A99} and \citet{C00} are plotted in
 Figure~\ref{fig9}.

 \section{Period-Luminosity Relations}  \label{PL}
 
 We have fit $B$ and $V$ period-luminosity $(P-L)$ relations of the form
 $M = \alpha \log P + \beta$ to the data for the fundamental mode and
 first overtone Cepheids, using all the Cepheids from Table~\ref{tbl6}
 with quality ``1'' photometry in the relevant passbands.
 Luminosity-weighted $\langle{B}\rangle_0$ and $\langle{V}\rangle_0$
 magnitudes have been used in these fits and no outliers have been
 excluded.  Coefficients from least squares and least absolute
 deviation fits are listed in Table~\ref{tbl9}.  The values of
 $\sigma$ listed for the least squares fits are the usual mean errors.
 Those listed for the least absolute deviation fits are a measure of
 the mean absolute deviation.  The period-luminosity relations are
 plotted in Figures~\ref{fig11},~\ref{fig12},~\ref{fig13},
 and~\ref{fig14}.

The least squares and least absolute deviation fits are almost
indistinguishable from one another and return coefficients which are
well within the uncertainties of the coefficients in the least squares
fits.  The scatter about the relationships, as measured by the
residuals ($\sigma$) from both methods, is similar. Some of the
scatter will, of course, be real -- reflecting the existence of a
finite width to the Cepheid instability strip in the HR diagram.  Some
of the scatter in the relationships may also be the result of
differential reddening across our four fields constituting our
northeast arm field.  Given the outlying location of our SMC field,
and the relatively small range in SMC reddening found by \citet{U99b},
the likelihood of a large range of reddenings for individual stars is
probably small. Nonetheless, differential reddening cannot be entirely
dismissed as a possible cause of scatter.

To check for the effect of differential reddening, we have calculated
the extinction-insensitive Wesenheit index, $W_{BFM}$, \citep{MF91} for
each of the Cepheids, where
\begin{eqnarray}
W_{BFM} & = & V - 3.24 \times (\bv)\,.
\end{eqnarray} 
The coefficients of fits to the $W_{BFM}$ versus $\log P$ diagrams are
given in Table~\ref{tbl10} and the relations are plotted in
Figure~\ref{fig15}. The scatter of the $W_{BFM}-\log P$ relation isn't
smaller than that for the extinction-corrected $\langle{V}\rangle -
\log P$ relations; indeed it is even larger for the first overtone
Cepheids, though the difference is not statistically
significant.  Nevertheless we are reassured that the scatter in the
$P-L$ relations for $B$ and $V$ is not attributable to differential
reddening across our field.

The coefficients of the $P-L$ relations derived from
our data can be compared with those derived from the photographic
$B$ survey of \citet{S92}. The $P-L$ relations from \citet{S92} are:
\begin{eqnarray}
\langle{B}\rangle_0 = -2.71(\pm 0.09)\log P_0 + 18.03(\pm 0.14) \,,\\
\langle{B}\rangle_0 = -2.98(\pm 0.17)\log P_1 + 17.35(\pm 0.14) \,,
\end{eqnarray}
\noindent for fundamental mode and first overtone mode Cepheids
respectively. These differ from those in Table~\ref{tbl9} in having
steeper slopes and fainter zero-points.

At $\log P = 0.8$, near $B$ = 15.8, the difference between the CCD and
photographic $P-L$ relations for fundamental mode Cepheids is about
0.1 mag.  At $\log P = 0.2$, near $B$ = 17.3, the $P-L$ relation
derived from the CCD data is about 0.22 mag brighter than that
obtained by \citet{S92}.  Of these differences, 0.04 mag is accounted
for by the different adopted reddening values in the two studies.  In
addition, the mean magnitudes given in \citet{S92} are magnitude
averages, rather than the luminosity-weighted magnitudes used here.
The effect which this has depends upon the lightcurve shape of the
variable under consideration.  For a typical fundamental mode Cepheid,
the luminosity-weighted magnitudes will be about 0.08 mag brighter
than those obtained from averaging magnitudes. These two differences
account for essentially all of the discrepancy in the two $P-L$
relations near $\log P$ = 0.8.  Together, however, they account for
only about half of the discrepancy near $\log P = 0.2$.

As will be discussed below, there may be a break in the slope of the
$P-L$ relation near a period of 2 days for fundamental mode Cepheids.
This break was not noted in \citet{S92}, although there is some
evidence for its existence in their $P-L$ relation. There are
different numbers of Cepheids with periods longer and shorter than two
days in the present study as compared to \citet{S92}. The stars of
period smaller than the break thus enter with somewhat different
weight into the calculation of the photographic and CCD $P-L$
relations.  This is responsible for a portion of the remaining
discrepancy at fainter magnitudes. Nonetheless, comparison of stars
common to the two studies suggests that, at $B \sim 17$, there is a
systematic difference of about 0.05 mag between the photographic and
CCD magnitude systems, with the latter being the brighter. We note
that there is some evidence for a field dependent magnitude error of
about 0.1 mag in the photographic photometry \citep{S92,G75}, so that
this zero-point shift may not perfectly apply across the entire
photographic field.

The comparison of the $B$ $P-L$ relations for the first overtone
Cepheids is more complicated, being influenced by the inclusion in
this study of significant numbers of variables not discovered in the
earlier photographic survey.  Part of the difference between the
photographic and CCD $P-L$ relations appears to be attributable to
this.  Again, however, comparison of the lightcurves of individual
variables indicates a systematic difference of 0.05 mag at $B$ $\sim$
17, in the sense that the CCD photometry is brighter.

\citet{C00} predict that, at Z = 0.004, the slope of the $B$ $P-L$
relation will be -2.49, for all fundamental mode Cepheids, and -2.71,
if limited to only those Cepheids having $\log P_0 < 1.5$.  For $V$,
the corresponding theoretical slopes in \citet{C00} are -2.60 and
-2.94.  The Cepheids in this study have periods less than $\log P_0$ =
1.5, but the $P-L$ relation slopes listed in Table~\ref{tbl9} are
slightly less steep than those predicted for such stars. As shown in
Figure~\ref{fig16}, a quadratic fit to the $V$ $P-L$ relation
describes our data better than a single linear $P-L$ relations.
However, because relatively few longer period Cepheids are included in
our sample, it is not possible to make a meaningful quantitative
comparison with the theoretical second order $P-L$ relationships given
in \citet{C00}.  The theoretical Wesenheit relation for fundamental
mode Cepheids in Table 7 of \citet{C00} has a slope of -3.65 $\pm$
0.02, which is very close to the observed value of -3.63 $\pm$ 0.12 in
Table~\ref{tbl10}.
 
The {\sc eros} group \citep{Ba99} reported the existence of a change
in the slope of the $P-L$ relation for fundamental mode Cepheids at
periods shorter than 2 days.  This steepening of the $P-L$ slope was
also noted by the {\sc ogle ii} group \citep{U99a}.  Our sample size is
smaller than those of the {\sc eros} and {\sc ogle ii} studies, making
the recognition of a change in slope difficult.  Nevertheless, when our
population of fundamental mode Cepheids is divided at a period of 2
days, it is evident that the two samples possess distinctly different
slopes.  This is shown graphically in Figures~\ref{fig17}
and~\ref{fig18}, and coefficients for the resultant fits are listed in
Table~\ref{tbl11}.

\citet{Ba99} note that a similar shift is not clearly seen in the
$P-L$ relation of the first overtone Cepheids.  To demonstrate this
they divide their first overtone population into samples at $P=1.4$
days, the period of a first overtone Cepheid which would correspond to
a 2 day fundamental mode period.  Although some steepening of their
$P-L$ relationship for first overtone mode Cepheids was present at
periods shorter than 1.4 day, the change in slope at that period was
not statistically significant.

In Table~\ref{tbl11} we list the coefficients we obtain fitting $P-L$
relations to our sample of first overtone mode Cepheids, considering
separately those with periods longer and shorter than 1.4 day.  The
overtone slopes, while they do show a tendency to steepen below a
period of 1.4 days, do not demonstrate as significant a difference as
is evident for the fundamental mode pulsators.  The difference in the
coefficients of the slope for periods above and below 1.4 days falls
within the limits expected from the uncertainties of the coefficients.
This is consistent with the findings of \citet{Ba99}.

\citet{Ba99} considered four possible explanations for the change in
the $P-L$ slope of the fundamental mode Cepheids: superposition of two
Cepheid populations of different ages and distances, mixing of
classical Cepheids and anomalous Cepheids, non-uniform filling of the
instability strip, and a thinning of the instability strip at short
periods. \citet{S92} had also noted the possibility that some of the
shorter period fundamental mode Cepheids might actually be anomalous
Cepheids.

\citet{Ba99} found no clear evidence for the superposition hypothesis,
particularly since they detected no significant change in the $P-L$
slope for the shortest period first overtone Cepheids.  They further
argued that the faint Cepheids were unlikely to be anomalous Cepheids
of the sort present in dwarf spheroidal systems because a known
metallicity dependence of the anomalous Cepheid $P-L$ relation
\citep{N94} would require metallicities much lower than are thought to
be common in the stellar population of the SMC.\@  We note, however, that
the dependence of the anomalous Cepheid $P-L$ relation upon
metallicity found by \citet{N94} is tied to an adopted
metallicity-luminosity relation for RR Lyrae stars.  The slopes of the
anomalous Cepheid $P-L$ relations in \citet{N94} are, however,
shallower than those reported in Table~\ref{tbl11} for fundamental
mode Cepheids with periods smaller than 2 days, which is an argument
against the anomalous Cepheid hypothesis.

There is some theoretical support for the idea that the lower end of
the Cepheid instability strip might not be completely filled.
Cepheids cross into the instability strip during the blue-loop phase
of their evolution, with the extent of the loop increasing with mass.
Low mass fundamental mode Cepheids are reasoned not to have as broad
an excursion into the instability strip, indeed eventually missing
portions of the strip, then missing the entire strip as the mass
continues to decrease.  The result is a truncation of the portion of
the instability strip that can be physically occupied by stars
pulsating in the fundamental mode.  This decreases the ``width'' of
the observed instability strip for fundamental mode Cepheids, with a
consequent shift in the $P-L$ relation.  The models of \citet{Ba98} and
\citet{A99} show that such an effect can produce a change in the slope
of the $P-L$ relation.

This hypothesis suffers from the lack of a similar observed change in
the first overtone $P-L$ relation, as noted by \citet{Ba99}.  The
region of the instability strip occupied by first overtone pulsators
is bluer than that occupied by the fundamental mode pulsators.  Stars
with blue-loops unable to extend into the fundamental portion of the
strip ought not to be able to extend into the overtone portion of the
strip either.  The result would be a depopulation of the overtone
sequence at a period corresponding to the luminosity at which the
effect is first noticed in the fundamental sequence.  A period of 2
days for the fundamental mode Cepheids would translate to the
aforementioned 1.4 days for overtone mode Cepheids. Thus, were this
explanation correct, we would expect the first overtone sequences to
be depopulated at periods shorter than 1.4 days. While the {\sc eros}
photometry \citep{Ba99}, our photometry, and the {\sc ogle ii} photometry
\citep{U99b} all indicate that there may be some steepening of the
$P-L$ slope for short period first overtone Cepheids, in each case the
effect is of doubtful statistical significance. Moreover, relatively
large numbers of first overtone mode Cepheids are present at periods
shorter than 1.4 days.  

\citet{A99} note that their models suggest that the change in slope
of the first overtone $P-L$ relation might be less pronounced, due to
a narrowing of the portion of the instability strip occupied by first
overtone mode pulsators with respect to that portion in which
fundamental mode pulsators reside.  However, \citet{A99} further
noted, though, that their models require that first overtone Cepheids
in the SMC with periods shorter than one day should be in the first
crossing of the instability strip.  Because first crossing stars
evolve quickly through the instability strip, they cannot explain the
relatively large number of short period first overtone Cepheids which
are actually observed, unless stars immersed in this phase of their
evolution somehow preferentially pulsate in the first overtone mode.

Our color-magnitude data for SMC Cepheids, and those from
\citet{U99b}, do indicate that there may be some change either in the
boundaries of the instability strip, or in the occupied region of the
instability strip, at lower luminosities.  Figures~\ref{fig7}
and~\ref{fig8} show that the redward limit of the region occupied by
fundamental mode pulsators shifts to bluer colors at
$\langle{V}\rangle_0 \ge 17$.  This shift is also noticeable as a
slight bend at the very faint end of the period-color relation for
fundamental mode Cepheids. The situation for the first overtone
pulsators is less clear, particularly because of the redder first
overtone pulsators seen in SC10 and SC11 but not within our field.  If
all else were equal, a blueward shift in the red boundary of the
instability strip would tend to remove the longer period Cepheids from
the sample at a given luminosity. By itself, this would not produce a
shift in the right direction to explain the observed change in $P-L$
slope.

\citet{U99a} applied an iterative least squares fitting routine to
fundamental mode Cepheids with periods above $\log P=0.4$, a cutoff
they selected to avoid the possible change in slope among the short
period Cepheids.  Their initial fit in $V$ yielded a shallower slope
(-2.57) for the $V$ $P-L$ relation in the SMC compared to that derived
from their LMC sample (-2.78).  On the basis of the more numerous
population of longer period ($\log P \ge 0.4$) fundamental mode
Cepheids in the LMC, and the smaller scatter about the LMC $V$ $P-L$
relation, \citet{U99a} decided to adopt the steeper LMC slope as the
effective SMC slope, and consequently re-fitted their SMC data.  In
Table~\ref{tbl12} we list both their original SMC fits, as well as the
$V$ fit obtained with the adopted, steeper LMC slope.  The $P-L$
relations listed in Table~\ref{tbl12} and previously quoted
coefficients are those from the {\sc ogle} website, and so include the
revisions made in 2000 \citep{U00}.  A direct comparison between our
relations and those from the {\sc eros} study \citep{Ba99} was not
made since they employ photometric bandpasses different than Johnson
$B$ and $V$.

The $P-L$ relations in Table~\ref{tbl12} may be compared with those
obtained for the fundamental mode Cepheids in our sample having
periods longer than 2 days (Table~\ref{tbl11}).  The least squares fit to our
sample gives a slope consistent with the \citet{U99a}
original fit to their $V$ data.  However, the uncertainty is large
enough to encompass the steeper slope of their adopted fit.  The
zero-points of the $P-L$ relations of our sample are, however,
slightly smaller than those found by \citet{U99a}.  That is
particularly so for their ``adopted" fit, but is true to a lesser
extent for their ``original" fit.

\section{Relative distance and depth of the northeast arm Cepheids}

There has been some debate regarding the depth of the SMC along the
line of sight and whether different regions of the SMC lie at
different mean distances.  \citet{CC86} used $BVRI$ observations of 63
Cepheids to conclude that we see the central bar of the SMC about edge
on, and that it has a depth of about 10 kpc.  They found the northeast
arm to be nearer to us than the central bar, and the southwest region
to be more distant, though with a superimposed nearer component.
\citet{M86} argued for a distance spread of some 30 kpc among SMC
Cepheids, while \citet{M88} used observations of 61 Cepheids to argue
that the SMC as a whole has a depth of some 20 kpc.  They interpreted
their data as showing that the northeastern arm Cepheids were some
10-15 kpc closer than those in the southern section.  However, they
found the Cepheids in the northeastern arm to be only slightly nearer
than those of the central bar.  In contrast, \citet{W87} concluded
from infrared observations of SMC Cepheids that the depth of the SMC
was smaller, and that the Cepheids generally fell within the circa 4
kpc tidal radius of the SMC.

\citet{HH89} investigated the distribution of a different stellar
population, using photometry of horizontal branch/red clump stars in
outlying northeastern and southwestern regions to conclude that the
older SMC population shows a relatively large line of sight depth.  In
their northeastern region, which is further from the central part of
the SMC than our field, they found a line of sight depth of 17 kpc on
average.

To provide further insight into this issue, we can compare the $P-L$
relations for our northeast arm field with those \citet{U99a}
obtained for the main body of the SMC, and with those which may be
derived from the Udalski et al.\ data for the more northeasterly of
the fields covered in their survey.

We have constructed $P-L$ relations for Cepheids in {\sc ogle ii} fields
SC10 and SC11 (Udalski et al.\ 1999b, using updated photometry in
Udalski et al.\ 2000) which, as we have noted earlier, are the {\sc
ogle ii} fields closest on the sky to our own.  These $P-L$ relations are
listed in Table~\ref{tbl13} and depicted in Figures~\ref{fig19}
through~\ref{fig22}.  For the fits to the fundamental mode Cepheids,
the slopes of the SC10--11 $P-L$ relations are steeper than those
obtained for our northeast arm field and the zero points are about
0.08 or 0.09 magnitudes fainter than those which we obtained
(Table~\ref{tbl9}), but the differences are significant only at the
one sigma level.  At $\log P = 0.4$, the $P-L$ relations for our
fundamental mode Cepheids predict $B$ and $V$ magnitudes which are
brighter by 0.04 mag or less than those predicted by the $P-L$
relations derived for the Cepheids in SC10 and SC11.  The $P-L$
relations for the first overtone mode Cepheids are somewhat more
discrepant, but this is mainly attributable to the faint, redder first
overtone Cepheids which, as we have noted, seem to be present in SC10
and SC11, but not within our field.  We conclude that our northeast
arm Cepheids can be at most only slightly closer than those in SC10
and SC11; i.e.\ there is no steep gradient in mean distance in going
along the northeast arm from the SC10 and SC11 fields to our more
outlying field.

The scatter about the mean $P-L$ relation for each field is
contributed by at least three separate effects: observational
uncertainty (including blending of images and differential reddening),
the finite color width of the $P-L$ relation, and the geometric depth
of the SMC in the region under consideration.  The scatter of the
individual Cepheids about the mean $P-L$ relation is smaller for our
dataset ($\sigma$ = 0.20 in $V$) than for the SC10 and SC11 dataset
($\sigma$ = 0.24).  However, it is unclear whether this signifies a
genuine difference between these fields.  The northeast arm field is
less crowded than SC10 and SC11, on average, and in deriving its $P-L$
relation we included only Cepheids with good lightcurves.

Correcting for the observed width of the Cepheid domain in the
color-magnitude diagram (see Figure~\ref{fig7}), the residual scatter
of the northeast arm Cepheids about the $P-L$ relations is about
$\sigma$ = 0.15, when results from both fundamental mode and first
overtone mode Cepheids are combined.  Taken at face value, this
corresponds to a depth along the line of sight of $\pm$ 4 kpc.  There
are, of course, outliers for which a distance greater than 4 kpc from
the mean would be determined, but it is also true that some of the
deduced range in distance is actually attributable to observational
error rather than a true distance spread.  It is difficult to decide
therefore the extremes in distance of the Cepheids in the northeast
arm field.  It is clear, however, that most Cepheids within this field
are located within $\pm$ 4 kpc of the mean distance, and that a very
large proportion are located within $\pm$ 6 kpc.

The differences between our $P-L$ relations for the northeast arm
field and those derived by \citet{U99a} from their entire SMC Cepheid
dataset are somewhat greater.  The least squares fit to the $V$ data
for our longer period fundamental mode Cepheids (Table~\ref{tbl11}),
predicts $\langle V \rangle_0 = 16.33 \pm 0.07$ at $\log P$ = 0.4,
whereas Udalski et al.'s (1999a, 2000) original fit to the full {\sc
ogle ii} dataset (Table~\ref{tbl12}) predicts $\langle V \rangle_0$ = 16.46
$\pm$0.02, and their adopted fit gives $\langle V \rangle_0$ = 16.53.
The corresponding predictions in the $B$ passband are $\langle B
\rangle_0$ = 16.78 $\pm$ 0.09 for our fit, and 16.84 $\pm$ 0.03 for
the original fit of \citet{U99a,U00}.  The differences for the first
overtone $P-L$ relations are larger, but, as noted earlier, they are
more difficult to interpret because of the apparent presence of some
types of first overtone Cepheids in the {\sc ogle ii} fields not found
within our northeast arm field.

Considering the respective uncertainties, these differences must be
regarded as of marginal significance.  Nonetheless, they are
consistent with the northeast arm being closer than the main body of
the SMC by about 4 kpc. The smaller scatter of the observations about
the mean $P-L$ relation for our northeast arm Cepheids compared to the
entire {\sc ogle ii} dataset is consistent with the existence of a
greater dispersion in distances among the entire {\sc ogle ii} Cepheid
population than among the Cepheids of the northeast arm.  However,
because the entire {\sc ogle ii} sample also includes some Cepheids which
suffer significant problems with crowding, photometric errors may also
contribute to that increased scatter.

\section{RR Lyrae Stars}

Two groups of RR Lyrae variables are apparent in Figure~\ref{fig4}.
Foreground field RR Lyrae stars have the periods and lightcurve
shapes typical of RR Lyrae stars, but their mean magnitudes scatter
between 15 and 18.  Then there are the actual SMC RR Lyrae stars,
clustered at $V \ge 19$.  The survey of SMC RR Lyrae stars within our
field is very incomplete, and their lightcurves are fairly noisy.
Even those SMC RR Lyrae labeled as having quality ``1'' lightcurves in
Table~\ref{tbl6} have poorer quality lightcurves than quality ``1'' SMC
Cepheids (but better lightcurves than the SMC RR Lyrae labeled
quality ``2''). The seven RRab stars in the SMC with quality ``1'' $V$
lightcurves have $\langle{V}\rangle$ = 19.39 $\pm$ 0.06.  The five
RRab stars with relatively good $B$ lightcurves have
$\langle{B}\rangle$ = 19.74 $\pm$ 0.08.  These values are likely to be
biased because the RR Lyrae with the better lightcurves also tend to
be among the brighter SMC RR Lyrae in the sample.  The mean value of
the luminosity-weighted $\langle{V}\rangle$ values for 11 probable SMC
RR Lyrae stars (including some with poorer lightcurves) is 19.43
$\pm$ 0.05.  The luminosity-weighted mean magnitude for the 9 SMC RR
Lyrae with usable values of $\langle{B}\rangle$ is 19.78 $\pm$ 0.06.
Smith et al.\ (1992) found a range in $\langle{B}\rangle$ among the RR
Lyrae stars of the northeast arm of the SMC, with a mean magnitude
weighted $B$ of 19.88 $\pm$ 0.04.  The corresponding magnitude
weighted mean $B$ magnitude for the present study is 19.83 $\pm$ 0.05,
in good agreement considering the uncertainties, even though the
search for RR Lyrae stars in the present study is more incomplete than
in the earlier photographic survey.

Our mean magnitudes for the RR Lyrae stars in the northeast arm field
are slightly brighter than the intensity weighted averages
\citet{WM88} found for RR Lyrae stars in the SMC cluster
\objectname{NGC 121}, $\langle{V}\rangle$ = 19.58 and
$\langle{B}\rangle$ = 19.91, or Graham's (1975) values for RR Lyrae
stars in the field around \objectname{NGC 121}, $\langle{V}\rangle$ =
19.57 and $\langle{B}\rangle$ = 19.95.  It is difficult to state to
what degree the incompleteness of the RR Lyrae sample in the present
survey is responsible for the difference.  However, because of the
possibility that our survey is biased toward the brighter RR Lyrae
stars, the differences with the Graham and Walker \& Mack studies
cannot be regarded as evidence that the RR Lyrae population in our
northeast arm field is nearer than that around \objectname{NGC 121}.

\section{Conclusions}

We observe a Cepheid population in the northeast arm which is in most
respects indistinguishable from that seen in prior studies of other
regions of the SMC.\@  The absence of the faint, redder first overtone
Cepheids found in SC10 and SC11 is a possible exception to that
conclusion.  To within the uncertainties, the slopes of the
period-luminosity relations we derive for the northeast arm field are
identical to those which have been obtained for more central regions
of the SMC.\@ The northeast arm Cepheids may be slightly nearer on
average than the Cepheids in the main body of the SMC, but the
difference is no greater than about 4 kpc.  The Cepheids in the
northeast arm region covered by our observations are, at most, only
slightly nearer than those in {\sc ogle ii} fields SC10 and SC11,
which lie between our field and the central area of the SMC.

Although the number of faint fundamental mode Cepheids within our
sample is relatively small, our data support the existence of a break
in the slope of the $P-L$ relation for fundamental mode Cepheids at
periods shorter than 2 days.  There is evidence for a similar break in
the $P-L$ relation for the first overtone mode Cepheids near a period
of 1.4 days, but in this case the evidence for the break is not
statistically significant.  The independent {\sc eros} and {\sc ogle ii}
studies also found marginal evidence for such a break in the slope of
the first overtone $P-L$ relation, which strengthens the case for its
reality, but not conclusively.

The first overtone mode and fundamental mode Cepheids occupy distinct
regions of the $V, (B-V)$ CMD.  There is some overlap between the
fundamental mode and first overtone mode Cepheid regions in the CMD,
but this overlap can in most cases be attributed to the observational
uncertainty of the photometry.  In the few cases in which the overlap
in color is too large to be explained in this fashion, the existence
of a few instances of unresolved red or blue companions to the
Cepheids provides a satisfactory alternative.  Below
$\langle{V}\rangle$ = 17, the red edge of the region of the CMD
occupied by fundamental mode pulsators shifts to bluer colors.  This
shift does not, by itself, provide an explanation for the break in the
slope of the $P-L$ relation at $P = 2$ days.

The period-color relations we obtain are in excellent agreement with
those found in analyses of Cepheids in the {\sc ogle ii} survey.
However, the observed period-color relation for fundamental mode
Cepheids is significantly bluer than the theoretical period-color
relation of \citet{A99} and is slightly bluer than that of
\citet{C00}.

\begin{acknowledgements}

We acknowledge use of the Digitized Sky Survey, which was produced at
STScI based on photographic data obtained using the UK Schmidt
Telescope, operated by the Royal Observatory Edinburgh.  The research
described in this paper was partially carried out by the Jet
Propulsion Laboratory, California Institute of Technology, under a
contract with NASA.  We thank Jason Lisle for his assistance in early
stages of the data reduction.  This work has been supported in part by
the National Science Foundation under grants AST-9317403 and
AST-9986943.

\end{acknowledgements}

\clearpage
\figurenum{1a} \figcaption[Sharpee.fig1a.ps]{Finding chart for newly
discovered variable stars (SSP numbers) in the northwest quadrant
(field 1).  East is at the bottom and north is at the
left. \label{fig1a}}

\figurenum{1b} \figcaption[Sharpee.fig1b.ps]{Finding chart for newly
discovered variable stars (SSP numbers) in the northeast quadrant
(field 2).  East is at the bottom and north is at the
left. \label{fig1b}}

\figurenum{1c} \figcaption[Sharpee.fig1c.ps]{Finding chart for newly
discovered variable stars (SSP numbers) in the southeast quadrant
(field 3).  East is at the bottom and north is at the
left. \label{fig1c}}

\figurenum{1d} \figcaption[Sharpee.fig1d.ps]{Finding chart for newly
discovered variable stars (SSP numbers) in the southwest quadrant
(field 4).  East is at the bottom and north is at the
left. \label{fig1d}}

\figurenum{2} \figcaption[Sharpee.fig2.ps]{Finding chart for the local
photometric standards. North is at the bottom and east is at the
left. \label{fig2}}

\figurenum{3} \figcaption[Sharpee.fig3.ps]{$B$ and $V$ light curves
for variable stars within the northeast arm field.  Variables observed
in both $B$ and $V$ are shown first.  Variables observed in only a
single band are included at the end of the figure.
\label{fig3}}  

\figurenum{4} \figcaption[Sharpee.fig4.ps]{Pulsating variable stars with
quality ``1''(see Table~\ref{tbl6}) light curves in both $B$ and $V$ in
the $\langle{V}\rangle$ versus $\log P$ diagram. Fundamental mode
Cepheids, first overtone mode Cepheids, and RR Lyrae stars are
identified. \label{fig4}}

\figurenum{5} \figcaption[Sharpee.fig5.ps]{Histograms over period for
fundamental mode and first overtone mode Cepheids reported in this
paper and in \protect{\citet{S92}} \label{fig5}}

\figurenum{6}
\figcaption[Sharpee.fig6.ps]{Histograms over period for quality ``1'' Cepheids
and for all Cepheids reported in this paper. \label{fig6}}

\figurenum{7} \figcaption[Sharpee.fig7.ps]{${\langle V \rangle}_o -
(\bv)_{mag,o}$ Color-magnitude diagram for fundamental mode (filled
triangles) and first overtone mode (open circles) Cepheids in the
northeast arm. Only those stars with quality ``1'' light curves in
both $B$ and $V$ are included.  \label{fig7}}

\figurenum{8} \figcaption[Sharpee.fig8.ps]{Same as Figure~\ref{fig7},
but including fundamental mode (filled squares) and first overtone
mode (open squares) Cepheids from fields SC10 and SC11 of
\protect{\citet{U99b}}. \label{fig8}}

\figurenum{9} \figcaption[Sharpee.fig9.ps]{Period-color relations for
fundamental mode Cepheids.  Shown are least squares fits to our
Cepheids with quality colors and to the {\sc ogle ii} dataset
\protect{\citep{U99b}}, as well as the theoretical period-color
relations of \protect{\citet{A99}} and \protect{\citet{C00}}. The
former two overlap. \label{fig9}}

\figurenum{10} \figcaption[Sharpee.fig10.ps]{Period-color relations
for first overtone mode Cepheids.  Least squares fits to our Cepheids
with quality colors and to the {\sc ogle ii} dataset
\protect{\citep{U99b}} are shown.
\label{fig10}}

\figurenum{11} \figcaption[Sharpee.fig11.ps]{$B$ period-luminosity
relation for fundamental mode Cepheids in the northeast arm
field. Least squares and least absolute deviation fits are
shown. \label{fig11}}

\figurenum{12} \figcaption[Sharpee.fig12.ps]{$B$ period-luminosity
relation for first overtone mode Cepheids in the northeast arm field.
The overlapping least squares and least absolute deviation fits are
shown. \label{fig12}}

\figurenum{13} \figcaption[Sharpee.fig13.ps]{$V$ period-luminosity
relation for fundamental mode Cepheids in the northeast arm field. The
overlapping least squares and least absolute deviation fits are
shown.\label{fig13}}

\figurenum{14} \figcaption[Sharpee.fig14.ps]{$V$ period-luminosity
relation for first overtone mode Cepheids in the northeast arm
field. The overlapping least squares and least absolute deviation fits
are shown. \label{fig14}}

\figurenum{15} \figcaption[Sharpee.fig15.ps]{Reddening insensitive $W_{BFM}$
versus $\log P$ relations for northeast arm Cepheids. Least squares fits
shown. \label{fig15}}

\figurenum{16} \figcaption[Sharpee.fig16.ps]{Second order fit to the $V$
period-luminosity relation for fundamental mode
Cepheids. \label{fig16}}

\figurenum{17} \figcaption[Sharpee.fig17.ps]{Change in slope of the
$B$ period-luminosity relation for fundamental mode Cepheids at $P =
2$ days.  Least squares and least absolute deviation fits are shown.
\label{fig17}}

\figurenum{18} \figcaption[Sharpee.fig18.ps]{Change in slope of the
$V$ period-luminosity relation for fundamental mode Cepheids at $P =
2$ days.  Least squares and least absolute deviation fits are
shown. \label{fig18}}

\figurenum{19} \figcaption[Sharpee.fig19.ps]{$B$ period-luminosity
relation for fundamental mode Cepheids in {\sc ogle ii} fields SC10
and SC11 \protect{\citep{U99b}}. Our linear least squares fit to the SC10 and
SC11 data is shown, as is the least squares fit to our northeast arm
sample. \label{fig19}}

\figurenum{20} \figcaption[Sharpee.fig20.ps]{$V$ period-luminosity
relation for fundamental mode Cepheids in {\sc ogle ii} fields SC10
and SC11. Fits shown are the same as in
Figure~\ref{fig19}. \label{fig20}}

\figurenum{21} \figcaption[Sharpee.fig21.ps]{$B$ period-luminosity
relation for first overtone mode Cepheids in {\sc ogle ii} fields SC10
and SC11. Fits shown are the same as in
Figure~\ref{fig19}. \label{fig21}}

\figurenum{22} \figcaption[Sharpee.fig22.ps]{$V$ period-luminosity
relation for first overtone mode Cepheids in {\sc ogle ii} fields SC10
and SC11. Fits shown are the same as in
Figure~\ref{fig19}. \label{fig22}}

\clearpage
\begin{deluxetable}{ccc} 
\tablewidth{0pc}
\tablecaption{Journal of Observations \label{tbl1}}
\tablehead{
\colhead{Heliocentric Julian Dates} & \colhead{Year} & \colhead{Months}
          }
\startdata
2448888 - 2448890 & 1992 & September\\
2448924 - 2448926 & 1992 & October \\
2449250 - 2449253 & 1993 & September \\
2449301 - 2449304 & 1993 & November \\ 
2449652 - 2449656 & 1994 & October \\
\enddata
\end{deluxetable}

\clearpage
\begin{deluxetable}{ccc}
\tablewidth{0pt}
\tablecaption{Coordinates of New Variables \label{tbl2}}
\tablehead{
\colhead{SSP} & \colhead{RA (J2000)} & 
\colhead{DEC (J2000)} 
}
\startdata
1 &  0:57:08.6 &-71:40:06 \\ 
2 &  0:57:15.1 &-71:49:26 \\ 
3 &  0:57:15.9 &-71:50:33 \\ 
4 &  0:57:27.5 &-71:40:57 \\ 
5 &  0:57:54.7 &-71:40:52 \\ 
6 &  0:57:58.7 &-71:36:37 \\ 
7 &  0:58:02.5 &-71:24:57 \\ 
8 &  0:58:13.4 &-71:26:55 \\ 
9 &  0:58:17.2 &-71:51:54 \\ 
10 &  0:58:24.7 &-71:42:38 \\ 
11 &  0:58:33.5 &-71:34:46 \\ 
12 &  0:58:43.3 &-71:42:38 \\ 
13 &  0:58:53.7 &-71:49:28 \\ 
14 &  0:58:55.1 &-71:39:03 \\ 
15 &  0:59:12.9 &-71:48:13 \\ 
16 &  0:59:13.4 &-71:34:46 \\ 
17 &  0:59:16.4 &-71:43:30 \\ 
18 &  0:59:31.5 &-71:51:40 \\ 
19 &  0:59:35.4 &-71:17:44 \\ 
20 &  1:00:02.3 &-71:42:08 \\ 
21 &  1:00:36.1 &-71:47:30 \\ 
22 &  1:00:54.3 &-71:44:44 \\ 
23 &  1:00:56.4 &-71:01:42 \\ 
24 &  1:00:59.8 &-71:12:59 \\ 
25 &  1:01:10.1 &-71:28:12 \\ 
26 &  1:01:16.8 &-71:48:41 \\ 
27 &  1:01:43.4 &-71:25:34 \\ 
28 &  1:01:56.7 &-71:16:31 \\ 
29 &  1:01:56.8 &-71:10:00 \\ 
30 &  1:02:21.6 &-71:29:49 \\ 
31 &  1:02:32.1 &-71:44:00 \\ 
32 &  1:02:42.6 &-71:08:22 \\ 
33 &  1:02:46.1 &-71:32:19 \\ 
34 &  1:02:50.8 &-71:43:45 \\ 
35 &  1:03:04.4 &-71:34:39 \\ 
36 &  1:03:14.9 &-71:44:53 \\ 
37 &  1:03:23.2 &-71:11:42 \\ 
38 &  1:03:48.3 &-71:22:12 \\ 
39 &  1:03:49.9 &-71:24:40 \\ 
40 &  1:03:51.1 &-71:42:02 \\ 
41 &  1:04:09.9 &-71:23:51 \\ 
42 &  1:04:20.5 &-71:40:38 \\ 
43 &  1:04:21.8 &-71:46:03 \\ 
44 &  1:04:28.7 &-71:04:19 \\ 
45 &  1:04:32.9 &-71:41:24 \\ 
46 &  1:04:38.2 &-71:40:08 \\ 
47 &  1:04:39.6 &-71:47:16 \\ 
48 &  1:04:45.1 &-71:53:41 \\ 
49 &  1:04:52.2 &-71:25:20 \\ 
50 &  1:05:01.2 &-71:43:34 \\ 
51 &  1:05:07.6 &-71:44:33 \\ 
52 &  1:05:28.1 &-71:40:11 \\ 
53 &  1:05:39.4 &-71:29:13 \\ 
54 &  1:05:49.7 &-71:46:56 \\ 
55 &  1:06:02.2 &-71:41:02 \\ 
56 &  1:06:08.5 &-71:49:18 \\ 
57 &  1:06:18.4 &-71:41:43 \\ 
58 &  1:06:30.0 &-71:08:04 \\ 
59 &  1:06:44.3 &-71:42:12 \\ 
60 &  1:06:49.0 &-71:29:22 \\ 
61 &  1:06:51.3 &-71:53:29 \\ 
62 &  1:06:53.3 &-71:37:05 \\ 
63 &  1:06:59.1 &-71:41:06 \\ 
64 &  1:07:03.5 &-71:39:15 \\ 
65 &  1:07:22.8 &-70:53:14 \\ 
66 &  1:07:42.2 &-71:42:56 \\ 
67 &  1:07:43.4 &-71:13:21 \\ 
68 &  1:07:50.6 &-71:33:03 \\ 
69 &  1:07:52.7 &-71:14:13 \\ 
70 &  1:07:53.9 &-71:50:18 \\ 
71 &  1:08:07.4 &-71:11:42 \\ 
72 &  1:08:08.7 &-71:17:24 \\ 
73 &  1:08:27.7 &-70:53:06 \\ 
74 &  1:08:32.9 &-71:29:13 \\ 
75 &  1:08:50.3 &-71:22:33 \\ 
76 &  1:08:59.8 &-71:34:14 \\ 
77 &  1:09:00.0 &-71:52:14 \\ 
78 &  1:09:02.8 &-70:55:13 \\ 
79 &  1:09:09.3 &-71:19:34 \\ 
80 &  1:09:10.2 &-71:30:28 \\ 
81 &  1:09:13.7 &-70:53:33 \\ 
82 &  1:09:15.6 &-70:55:28 \\ 
83 &  1:09:22.6 &-71:14:39 \\ 
84 &  1:09:25.0 &-71:18:11 \\ 
\enddata
\end{deluxetable}

\clearpage

\begin{deluxetable}{ccccrrr}
\tablewidth{0pt}
\tablecaption{Photometry of Local Standard Stars \label{tbl3}}
\tablehead{
\colhead{ID} & \colhead{RA (J2000)}      &
\colhead{DEC (J2000)} & \colhead{$V$} &
\multicolumn{1}{c}{$B-V$} & \multicolumn{1}{c}{$U-B$} &
\multicolumn{1}{c}{$V-I$}
}
\startdata
A & 1:05:59.9 & -71:26:11 & 13.173 & 0.628 &\nodata& 0.714 \\
  &           &           & 13.176 & 0.627 & 0.098 & 0.716 \\
B & 1:04:43.2 & -71:25:56 & 12.706 & 0.770 &\nodata& 0.819 \\
  &           &           & 12.706 & 0.774 & 0.386 & 0.817 \\
C & 1:05:53.4 & -71:22:39 & 14.074 & 0.760 &\nodata& 0.869 \\
  &           &           & 14.073 & 0.755 & 0.199 & 0.865 \\
  &           &           & 14.082 & 0.744 & 0.188 & 0.874 \\
D & 1:05:57.8 & -71:19:14 & 12.456 & 0.040 &\nodata& 0.114 \\
  &           &           & 12.457 & 0.040 & -0.121& 0.116 \\
E & 1:06:08.9 & -71:17:08 & 13.817 & 0.746 & 0.394 & 0.773 \\
  &           &           & 13.821 & 0.742 &\nodata& 0.771 \\
  &           &           & 13.825 & 0.740 & 0.377 & 0.778 \\
F & 1:05:06.6 & -71:18:06 & 13.422 & 0.643 & 0.180 & 0.698 \\
  &           &           & 13.435 & 0.651 &\nodata& 0.691 \\
G & 1:04:41.6 & -71:16:46 & 12.880 & 0.426 &\nodata& 0.540 \\
  &           &           & 12.879 & 0.422 & -0.096& 0.539 \\
H & 1:04:10.0 & -71:17:13 & 14.833 & 0.688 & 0.178 & 0.727 \\
  &           &           & 14.852 & 0.675 &\nodata& 0.755 \\
  &           &           & 14.846 & 0.679 & 0.171 & 0.742 \\
I & 1:03:53.8 & -71:18:54 & 15.076 & -0.240& -0.931& -0.247\\
  &           &           & 15.070 & -0.237&\nodata& -0.263\\
  &           &           & 15.066 & -0.233& -0.947& -0.266\\
J & 1:04:44.3 & -71:20:29 & 15.939 & 0.599 &\nodata& 0.734 \\
  &           &           & 15.934 & 0.613 & 0.071 & 0.710 \\
  &           &           & 15.951 & 0.606 & 0.051 & 0.745 \\
K & 1:04:53.3 & -71:19:59 & 15.331 & 0.757 & 0.318 & 0.828 \\
  &           &           & 15.335 & 0.760 &\nodata& 0.835 \\
  &           &           & 15.330 & 0.763 & 0.325 & 0.826 \\
L & 1:05:51.7 & -71:18:16 & 14.807 & 0.711 & 0.203 & 0.822 \\
  &           &           & 14.800 & 0.719 & 0.214 & 0.821 \\
  &           &           & 14.805 & 0.714 &\nodata& 0.818 \\
M & 1:04:13.5 & -71:25:16 & 15.092 & 0.836 & 0.386 & 0.986 \\
  &           &           & 15.087 & 0.824 &\nodata& 0.962 \\
  &           &           & 15.077 & 0.830 & 0.360 & 0.951 \\
N & 1:04:42.3 & -71:25:08 & 15.754 & -0.225& -0.894& -0.255\\
  &           &           & 15.744 & -0.218& -0.871& -0.256\\
  &           &           & 15.754 & -0.220&\nodata& -0.238\\
O & 1:05:56.2 & -71:20:25 & 16.829 & 1.486 & 0.621 & 1.443 \\
  &           &           & 16.830 & 1.522 &\nodata& 1.421 \\
  &           &           & 16.852 & 1.509 & 2.516 & 1.462 \\
P & 1:05:10.5 & -71:19:04 & 16.497 & 1.230 &\nodata& 1.300 \\
  &           &           & 16.483 & 1.244 & 0.898 & 1.298 \\
  &           &           & 16.483 & 1.219 & 0.949 & 1.282 \\
Q & 1:05:16.1 & -71:19:22 & 16.722 & 0.969 & 0.545 & 1.027 \\
  &           &           & 16.749 & 0.967 &\nodata& 1.052 \\
  &           &           & 16.747 & 0.933 & 0.430 & 1.029 \\
R & 1:04:52.0 & -71:25:39 & 16.909 & 0.967 &\nodata& 1.062 \\
  &           &           & 16.885 & 0.943 & 0.690 & 1.045 \\
  &           &           & 16.891 & 0.959 & 0.480 & 1.071 \\
S & 1:04:04.7 & -71:28:23 & 16.853 & 0.557 &\nodata& 0.824 \\
  &           &           & 16.846 & 0.551 & -0.011& 0.716 \\
  &           &           & 16.825 & 0.555 & 0.127 & 0.774 \\
T & 1:04:05.3 & -71:27:37 & 15.641 & 0.689 & 0.237 & 0.727 \\
  &           &           & 15.646 & 0.686 &\nodata& 0.758 \\
  &           &           & 15.640 & 0.686 & 0.211 & 0.751 \\
U & 1:05:31.4 & -71:21:32 & 15.688 & 0.026 &\nodata& 0.022 \\
  &           &           & 15.690 & 0.033 & -0.065& 0.025 \\
  &           &           & 15.689 & 0.027 & -0.038& 0.026 \\
\enddata
\end{deluxetable}

\clearpage

\begin{deluxetable}{cccc} 
\tablewidth{0pc}
\tablecaption{Comparison of Photometry \label{tbl4}}
\tablehead{
\colhead{Reference} & \colhead{n} & \colhead{$\Delta\,V$} & 
\colhead{$\Delta\,B$}
          }
\startdata
Harris & 12 & $0.021\pm0.010$ & $0.013\pm0.014$ \\
Da Costa et al. & 4 & \nodata & $-0.003\pm0.021$ \\
\enddata
\end{deluxetable}

\clearpage

\begin{deluxetable}{ccccrr} 
\tablewidth{0pc}
\tablecaption{Photometry of Individual Variable Stars \label{tbl5}}
\tablehead{
\colhead{Star ID}  &\colhead{Field} & \colhead{Bandpass} & \colhead{HJD} & 
\colhead{Mag} & 
\colhead{Error}
          }
\startdata
HV214   &    2 & B & 2448889.6461 & 16.588 & 0.007 \\
HV214   &    2 & B & 2448889.7790 & 16.597 & 0.007 \\
HV214   &    2 & B & 2448890.6559 & 16.797 & 0.008 \\
HV214   &    2 & B & 2448890.7980 & 16.811 & 0.007 \\
HV214   &    2 & B & 2448924.5873 & 16.848 & 0.009 \\
HV214   &    2 & B & 2448924.8235 & 16.790 & 0.009 \\
HV214   &    2 & B & 2448925.5839 & 15.451 & 0.004 \\
HV214   &    2 & B & 2448925.7343 & 15.535 & 0.004 \\
HV214   &    2 & B & 2448926.5962 & 16.156 & 0.006 \\
HV214   &    2 & B & 2448926.7483 & 16.242 & 0.006 \\
HV214   &    2 & B & 2449252.7222 & 17.056 & 0.008 \\ 
HV214   &    2 & B & 2449253.6185 & 15.305 & 0.003 \\ 
HV214   &    2 & B & 2449253.7711 & 15.538 & 0.003 \\ 
HV214   &    2 & B & 2449301.5958 & 16.553 & 0.005 \\ 
HV214   &    2 & B & 2449301.7486 & 16.560 & 0.006 \\
HV214   &    2 & B & 2449302.6621 & 16.798 & 0.007 \\
HV214   &    2 & B & 2449303.6598 & 15.969 & 0.004 \\
HV214   &    2 & B & 2449304.6476 & 15.890 & 0.004 \\
HV214   &    2 & B & 2449304.8149 & 15.993 & 0.004 \\
HV214   &    2 & B & 2449653.5560 & 15.846 & 0.004 \\
HV214   &    2 & B & 2449653.7614 & 15.960 & 0.005 \\
HV214   &    2 & B & 2449656.6026 & 16.839 & 0.008 \\
HV214   &    2 & V & 2448888.6283 & 15.767 & 0.005 \\
HV214   &    2 & V & 2448888.7495 & 15.691 & 0.005 \\ 
HV214   &    2 & V & 2448888.8414 & 15.781 & 0.005 \\ 
HV214   &    2 & V & 2448889.6633 & 16.003 & 0.004 \\
HV214   &    2 & V & 2448889.7967 & 16.038 & 0.005 \\
HV214   &    2 & V & 2448890.6714 & 16.154 & 0.005 \\
HV214   &    2 & V & 2448924.6517 & 16.212 & 0.005 \\
HV214   &    2 & V & 2448924.8426 & 16.301 & 0.008 \\ 
HV214   &    2 & V & 2448925.5980 & 15.154 & 0.003 \\
HV214   &    2 & V & 2448925.7532 & 15.266 & 0.003 \\ 
HV214   &    2 & V & 2448926.5716 & 15.846 & 0.005 \\ 
HV214   &    2 & V & 2448926.7215 & 15.791 & 0.004 \\
HV214   &    2 & V & 2449253.6007 & 14.985 & 0.003 \\
HV214   &    2 & V & 2449253.7889 & 15.220 & 0.003 \\
HV214   &    2 & V & 2449301.6124 & 15.965 & 0.004 \\
HV214   &    2 & V & 2449302.6453 & 16.201 & 0.006 \\
HV214   &    2 & V & 2449302.7927 & 16.199 & 0.005 \\
HV214   &    2 & V & 2449303.6426 & 15.652 & 0.004 \\
HV214   &    2 & V & 2449303.7860 & 15.345 & 0.003 \\
HV214   &    2 & V & 2449304.7977 & 15.585 & 0.004 \\
HV214   &    2 & V & 2449652.7206 & 15.533 & 0.005 \\ 
HV214   &    2 & V & 2449653.5739 & 15.451 & 0.004 \\
HV214   &    2 & V & 2449653.7438 & 15.533 & 0.004 \\
HV214   &    2 & V & 2449656.6205 & 16.138 & 0.005 \\
\enddata
\tablecomments{The complete version of this table is in the electronic
edition of the Journal.  The printed edition contains only a sample.}
\end{deluxetable}

\begin{deluxetable}{rlrrrr@{\hspace{1.25pc}}r@{\hspace{2pc}}cc}
\tablewidth{0pt}
\tablecaption{Attributes of Observed Variables \label{tbl6}}
\tablehead{
\multicolumn{1}{c}{Num} & 
\multicolumn{1}{c}{Type} & 
\multicolumn{1}{c}{P (days)} & 
\multicolumn{1}{c}{$\langle B \rangle$} & 
\multicolumn{1}{c}{$\langle V \rangle$} & 
\multicolumn{1}{c}{$\langle B \rangle - \langle V \rangle$} &
 \multicolumn{1}{c}{$(B-V)_{mag}$} & 
\multicolumn{1}{c}{Qual} & 
\multicolumn{1}{c}{Notes} 
}
\startdata
\sidehead{HV} 
 214 & F & 4.20494  &    16.232 &    15.775 & 0.457 &     0.498 &2 & 
 \\
 841 & F & 1.90951  &    17.010 &    16.696 & 0.314 &     0.347 &1 & 
 \\
 856 & F & 6.07682  &    15.936 &    15.139 & 0.797 &     0.810 &2 & 
 \\
 1717 & F & 1.87023  &    17.117 &    16.766 & 0.351 &     0.394 &1 & 
 \\
 1718 & F & 4.04534  &    16.439 &    15.954 & 0.485 &     0.497 &2 & 
 \\
 1727 & F & 2.22099  &    16.636 &    16.362 & 0.274 &     0.300 &1(V) & 
 \\
 1732 & F & 1.75459  &    17.158 &    16.844 & 0.314 &     0.323 &1(V) & 
 \\
 1755 & O & 1.50575  &    16.837 &    16.525 & 0.312 &     0.322 &1 & 
 \\
 1759 & F & 1.73444  &    17.156 &    16.758 & 0.398 &     0.404 &1 & 
 \\
 1779 & O & 1.78335  &    16.532 &    16.144 & 0.388 &     0.398 &1 & 
 \\
 1788 & O & 3.47885  &    15.762 &    15.338 & 0.424 &     0.428 &1 & 
 \\
 1793 & F & 4.18158  &    16.549 &    16.021 & 0.528 &     0.558 &2 & 
 \\
 1807 & F & 4.08858  &    16.569 &    16.067 & 0.502 &     0.523 &2 & 
 \\
 1855 & F & 6.83922  &    15.922 &    15.361 & 0.561 &     0.579 &1 & 
 \\
 1860 & O & 1.98478  &    16.345 &    16.001 & 0.344 &     0.349 &1 & 
 \\
 1863 & RRab & 0.51327  &    18.162 &    17.733 & 0.429 &     0.430 &2 & 
\tablenotemark{(a)}\\
 1869 & F & 2.46485  &    17.381 &    16.849 & 0.532 &     0.541 &1 & 
 \\
 1871 & F & 1.30090  &    17.515 &    17.175 & 0.340 &     0.380 &1 & 
 \\
 1883 & O & 2.01315  &    16.536 &    16.102 & 0.434 &     0.444 &1 & 
 \\
 1890 & F & 1.77851  &    17.215 &    16.825 & 0.390 &     0.428 &1 & 
 \\
 1892 & F & 5.65497  &    16.292 &    15.700 & 0.592 &     0.601 &2 & 
 \\
 1897 & O & 1.24131  &    17.089 &    16.738 & 0.351 &     0.360 &1(V) & 
 \\
 1898 & F & 3.01818  &    16.434 &    16.141 & 0.293 &     0.296 &2 & 
\tablenotemark{(b)}\\
 1901 & F & 3.57357  &    16.900 &    16.323 & 0.577 &     0.591 &1 & 
 \\
 1906 & F & 3.06560  &    16.759 &    16.230 & 0.529 &     0.551 &2 & 
 \\
 1907 & F & 1.64318  &    17.243 &    16.873 & 0.370 &     0.420 &1(B) & 
 \\
 1919 & F & 1.71874  &    17.501 &    17.071 & 0.430 &     0.478 &1 & 
 \\
 1929 & F & 5.58692  &    15.958 &    15.520 & 0.438 &     0.442 &1 & 
 \\
 1934 & F & 4.87545  &    16.294 &    15.758 & 0.536 &     0.560 &1 & 
 \\
 1945 & F & 6.47378  &    15.936 &    15.382 & 0.554 &     0.579 &1 & 
 \\ 
 1950 & F & 7.98574  &    15.752 &    15.108 & 0.644 &     0.655 &1(B) & 
 \\
 1958 & F & 1.33443  &    17.653 &    17.256 & 0.397 &     0.460 &1 & 
 \\
 1970 & F & 3.33609  &    16.630 &    16.195 & 0.435 &     0.484 &1(V) & 
 \\
 1976 & F & 1.90304  &    17.386 &    17.018 & 0.368 &     0.425 &1(B) & 
 \\
 1982 & F & 5.22421  &    16.259 &    15.659 & 0.600 &     0.618 &2 & 
 \\
 1983 & F & 3.43840  &    16.255 &    15.824 & 0.431 &     0.459 &1 & 
 \\
 1987 & F & 3.13062  &    16.671 &    16.167 & 0.504 &     0.522 &1 & 
 \\
 1989 & O & 1.08200  &    17.014 &    16.627 & 0.387 &     0.398 &1 & 
 \\
 1999 & F & 1.96777  &    17.351 &    16.905 & 0.446 &     0.468 &1 & 
 \\
 2002 & F & 2.34646  &    16.966 &    16.543 & 0.423 &     0.477 &1 & 
 \\
 2003 & F & 5.05307  &    16.571 &    15.932 & 0.639 &     0.646 &1 & 
 \\
 2013 & F & 2.85747  &    16.756 &    16.321 & 0.435 &     0.486 &1 & 
 \\
 2014 & F & 2.20378  &    16.803 &    16.412 & 0.391 &     0.445 &1 & 
 \\
 2015 & F & 2.87409  &    17.091 &    16.549 & 0.542 &     0.565 &1 & 
 \\
 2017 & F &11.40440  &    15.457 &    14.749 & 0.708 &     0.718 &2 & 
 \\
 2043 & RRab & 0.60337  &    15.593 &    15.220 & 0.373 &     0.389 &1 & 
 \\
 2045 & F & 2.84686  &    16.827 &    16.284 & 0.543 &     0.574 &1 & 
 \\
 2053 & F & 3.21897  &    16.805 &    16.332 & 0.473 &     0.499 &1 & 
 \\
 2054 & F & 7.16373  &    15.799 &    15.212 & 0.587 &     0.597 &2 & 
 \\
 2055 & F & 3.54469  &    16.974 &    16.363 & 0.611 &     0.615 &2 & 
 \\
 2057 & F & 1.90899  &    17.202 &    16.809 & 0.393 &     0.447 &1 & 
 \\
 2059 & F & 2.85985  &    16.739 &    16.328 & 0.411 &     0.422 &1(V) & 
 \\
 2063 & F &11.17360  &    15.425 &    14.777 & 0.648 &     0.672 &2 & 
 \\
 2068 & O & 1.80701  &    16.863 &    16.438 & 0.425 &     0.436 &1 & 
 \\
 2074 & O & 2.19972  &    16.487 &    16.120 & 0.367 &     0.378 &1 & 
 \\
 2076 & F & 2.49908  &    16.706 &    16.269 & 0.437 &     0.487 &1 & 
 \\
 2078 & F & 4.70610  &    16.330 &    15.801 & 0.529 &     0.559 &1 & 
 \\
 2085 & F & 3.71630  &    16.837 &    16.287 & 0.550 &     0.568 &1 & 
 \\
 2088 & F &14.57880  &    15.533 &    14.754 & 0.779 &     0.807 &2 & 
 \\
 2102 & RRab & 0.51404  &    16.162 &    15.863 & 0.299 &     0.339 &1 & 
 \\
 2107 & F & 2.41154  &    16.885 &    16.439 & 0.446 &     0.482 &1 & 
 \\
 10364 & F & 1.50304  &    17.870 &    17.238 & 0.632 &     0.683 &1(V) & 
 \\
 10365 & O & 1.25811  &    17.202 &    16.840 & 0.362 &     0.377 &1 & 
 \\
 10367-1 & F & 1.54812  &    17.488 &    17.109 & 0.379 &     0.427 &1 & 
 \\
 10367-2 & F & 2.76831  &    17.194 &    16.680 & 0.514 &     0.531 &1 & 
 \\
 10373 & F & 3.07538  &    16.940 &    16.362 & 0.578 &     0.627 &2 & 
 \\
 10375 & F & 5.94034  &    16.402 &    15.783 & 0.619 &     0.625 &1 & 
 \\
 11190 & O & 1.36617  &    17.030 &    16.670 & 0.360 &     0.370 &1(B) & 
 \\
 11191 & F & 1.96643  &    16.904 &    16.497 & 0.407 &     0.445 &1 & 
 \\
 11192 & F & 2.35403  &    16.964 &    16.570 & 0.394 &     0.423 &2 & 
 \\
 11197 & F? & 1.07507  &    17.109 &    16.713 & 0.396 &     0.409 &2 & 
 \\
 11199 & F & 2.09523  &    17.266 &    16.874 & 0.392 &     0.408 &1 & 
 \\
 11206 & F & 3.39809  &    16.691 &    16.195 & 0.496 &     0.507 &1 & 
 \\
 11209 & O & 1.92843  &    16.641 &    16.238 & 0.403 &     0.413 &1 & 
 \\
 11375 & O & 1.11490  &    17.134 & \nodata &  \nodata & \nodata &1 & 
 \\
 11395 & F & 1.29710  &    18.081 &    17.621 & 0.460 &     0.478 &1 & 
 \\
 11449 & O & 1.31566  &    17.057 &    16.686 & 0.371 &     0.377 &1 & 
 \\
 11457 & O & 2.27465  &    16.619 &    16.234 & 0.385 &     0.391 &1 & 
 \\
 11463 & F & 1.75536  &    17.625 &    17.176 & 0.449 &     0.473 &1 & 
 \\
 11485 & F & 1.20924  &    17.829 &    17.449 & 0.380 &     0.421 &1 & 
 \\
 11500 & O & 1.78969  &    16.620 &    16.244 & 0.376 &     0.385 &1 & 
 \\
 11501 & F & 3.57905  &    16.914 &    16.306 & 0.608 &     0.614 &2 & 
 \\
 11502 & ? & 3.27030  &    16.410 &    15.941 & 0.469 &     0.473 &2 & 
\tablenotemark{(c)}\\
 12942 & F & 3.89314  &    16.426 &    15.890 & 0.536 &     0.554 &2 & 
 \\
 12943 & F & 3.71538  &    16.643 &    16.090 & 0.553 &     0.572 &1 & 
 \\
 12949 & RRab & 0.47498  &    17.592 &    17.286 & 0.306 &     0.353 &1 & 
 \\
 13020 & O & 0.98915  &    17.424 &    17.111 & 0.313 &     0.326 &2 & 
 \\
 13021 & O & 1.81468  &    16.734 &    16.320 & 0.414 &     0.420 &1 & 
 \\
\sidehead{NGC362V}
 3 & RRab & 0.47324  &    15.836 &    15.533 & 0.303 &     0.356 &1 & 
 \\
 8 & F & 3.86046  &    16.355 &    15.896 & 0.459 &     0.514 &1(B) & 
 \\
 \sidehead{[SSB92]}  
 4 & O & 1.59263  &    17.114 &    16.649 & 0.465 &     0.478 &1 & 
 \\
 5 & F & 2.50085  &    17.144 &    16.657 & 0.487 &     0.509 &1 & 
 \\
 6 & RRab & 0.60537  &    19.689 &    19.240 & 0.449 &     0.474 &2 & 
 \\
 7 & F & 1.02203  &    18.130 &    17.751 & 0.379 &     0.394 &2 & 
 \\
 9 & RRab & 0.58500  & \nodata &     18.833 & \nodata & \nodata &2 & 
 \\
 10 & RRab & 0.59463  &    19.939 &    19.535 & 0.404 &     0.417 &2 & 
 \\
 11 & F & 1.07980  &    18.182 &    17.781 & 0.401 &     0.414 &2 & 
 \\
 12 & F & 1.33366  &    17.474 &    17.157 & 0.317 &     0.365 &1 & 
 \\
 14 & F & 1.72298  &    17.736 &    17.233 & 0.503 &     0.521 &1 & 
 \\
 18 & E & 0.31872  &    18.528 & \nodata &  \nodata & \nodata &2 & 
\tablenotemark{(d)}\\
 19 & E & 0.43382  &    18.422 &    18.343 & 0.079 &     0.078 &2 & 
 \\
 20 & F & 1.49082  &    17.616 &    17.192 & 0.424 &     0.452 &1 & 
 \\
 24 & F & 1.43056  & \nodata &     17.320 & \nodata & \nodata &2 & 
 \\
 25 & RRab? & 0.52774  & \nodata &     19.531 & \nodata & \nodata &2 & 
 \\
 26 & RRab & 0.44636  &    17.538 &    17.188 & 0.350 &     0.401 &1 & 
 \\
 27 & F & 1.50863  &    17.604 &    17.020 & 0.584 &     0.608 &1 & 
 \\
 29 & RRab & 0.47084  &    18.135 &    17.856 & 0.279 &     0.335 &1 & 
 \\
 31 & O & 0.58038  &    18.137 &    17.774 & 0.363 &     0.371 &1 & 
 \\
 32 & F & 2.26841  &    17.123 &    16.742 & 0.381 &     0.427 &1 & 
 \\
 34 & O & 1.93682  &    16.663 &    16.233 & 0.430 &     0.439 &1 & 
 \\
 36 & F & 2.11420  &    17.505 &    16.989 & 0.516 &     0.548 &1 & 
 \\
 37 & RRab & 0.59659  &    19.884 &    19.507 & 0.377 &     0.393 &1 & 
 \\
 38 & F & 1.63004  &    17.576 &    17.234 & 0.342 &     0.393 &2 & 
 \\
 42 & F & 1.33152  &    17.814 &    17.449 & 0.365 &     0.405 &1(B) & 
 \\
 45 & RRab & 0.52640  &    19.702 &    19.451 & 0.251 &     0.266 &1(V) & 
 \\
 46 & F? & 2.55475  &    17.206 &    16.671 & 0.535 &     0.545 &2 & 
 \\
 48 & F & 2.29939  &    17.395 & \nodata &  \nodata & \nodata &1 & 
 \\
 50 & F & 0.88530  &    18.653 &    18.380 & 0.273 &     0.308 &1 & 
 \\
 53 & O & 0.82857  &    17.639 &    17.335 & 0.304 &     0.311 &1 & 
 \\
 55 & E & 1.35994  &    18.666 &    18.673 &-0.007 &    -0.000 &1 & 
 \\
 57 & O & 0.72152  &    17.821 &    17.494 & 0.327 &     0.341 &1 & 
 \\
 59 & O & 0.86133  &    17.737 &    17.392 & 0.345 &     0.359 &1 & 
 \\
 61 & ? & 0.51718  &    18.757 &    18.122 & 0.635 &     0.635 &2 & 
\tablenotemark{(e)}\\
 62 & O & 0.78903  &    17.837 &    17.341 & 0.496 &     0.502 &1(V) & 
 \\
 65 & F & 2.00666  &    17.520 &    17.015 & 0.505 &     0.534 &1 & 
 \\
 66 & O & 1.19755  &    17.244 &    16.891 & 0.353 &     0.363 &1 & 
 \\
 68 & F & 2.04065  &    17.221 &    16.772 & 0.449 &     0.483 &1(V) & 
 \\
 70 & F & 5.65365  &    16.284 &    15.702 & 0.582 &     0.593 &1 & 
 \\
 72 & O & 0.91060  &    17.761 &    17.413 & 0.348 &     0.361 &1 & 
 \\
 73 & O & 1.26752  &    17.108 &    16.742 & 0.366 &     0.377 &1 & 
 \\
 74 & F & 1.52872  &    17.394 &    17.044 & 0.350 &     0.383 &1 & 
 \\
 75 & F & 1.52205  &    17.595 &    17.116 & 0.479 &     0.493 &2 & 
 \\
 77 & O & 1.35339  &    17.026 &    16.672 & 0.354 &     0.365 &1 & 
 \\
 80 & F & 1.28913  &    17.588 &    17.194 & 0.394 &     0.396 &1 & 
 \\
 84 & O & 0.76310  &    17.672 &    17.392 & 0.280 &     0.295 &1 & 
 \\
 87 & O & 0.88342  &    17.438 &    17.071 & 0.367 &     0.384 &1 & 
 \\
 88 & O & 0.81100  &    17.744 &    17.380 & 0.364 &     0.379 &1 & 
 \\
 89 & RRab & 0.45418  &    17.128 &    16.880 & 0.248 &     0.290 &1 & 
 \\
 90 & O & 2.71711  &    16.377 &    15.901 & 0.476 &     0.481 &1 & 
 \\
 92 & O & 0.84835  &    17.760 &    17.405 & 0.355 &     0.368 &1 & 
 \\
 95 & F & 2.61612  &    17.317 &    16.735 & 0.582 &     0.594 &1 & 
 \\
 96 & F & 1.34841  &    17.990 &    17.540 & 0.450 &     0.473 &1(V) & 
 \\
 103 & F & 1.54345  &    17.514 &    17.156 & 0.358 &     0.378 &2 & 
 \\
 105 & RRab & 0.59722  &    19.465 &    19.202 & 0.263 &     0.295 &1 & 
 \\
 107 & F & 6.49091  &    15.833 &    15.286 & 0.547 &     0.571 &1 & 
 \\
 110 & O & 2.06922  &    16.808 &    16.395 & 0.413 &     0.420 &1 & 
 \\
 111 & F & 1.83433  &    17.089 &    16.740 & 0.349 &     0.376 &2 & 
 \\
 113 & F & 1.08116  &    18.290 &    17.945 & 0.345 &     0.390 &1 & 
 \\
 114 & O & 0.97918  &    17.192 &    16.918 & 0.274 &     0.285 &2 & 
 \\
 117 & F & 3.09329  &    16.987 &    16.452 & 0.535 &     0.547 &1 & 
 \\
 124 & O & 0.86375  &    17.041 &    16.925 & 0.116 &     0.116 &2 & 
\tablenotemark{(f)}\\
 128 & O & 0.76036  &    18.099 &    17.735 & 0.364 &     0.377 &1 & 
 \\
 133 & F & 1.04962  &    18.048 &    17.710 & 0.338 &     0.368 &1 & 
 \\
\sidehead{SSP}
 1 & F & 1.87056  &    17.074 &    16.746 & 0.328 &     0.379 &1 & 
 \\
 2 & F & 2.23245  &    17.431 &    16.825 & 0.606 &     0.595 &2 & 
 \\
 3 & O & 1.17628  &    17.155 &    16.840 & 0.315 &     0.325 &1 & 
 \\
 4 & E & 2.30000  &    17.408 &    17.390 & 0.018 &     0.024 &2 & 
 \\
 5 & O & 1.16093  &    17.350 &    16.992 & 0.358 &     0.368 &2 & 
 \\
 6 & O & 1.29270  &    16.975 &    16.660 & 0.315 &     0.325 &1 & 
 \\
 7 & O & 0.93102  &    17.492 &    17.105 & 0.387 &     0.395 &1 & 
 \\
 8 & O & 2.07345  &    16.298 &    15.936 & 0.362 &     0.367 &1 &
 \\ 
 9 & O & 2.29499  &    16.439 &    16.044 & 0.395 &     0.401 &1 & 
 \\
 10 & F? & 1.70068  &    17.285 &    16.851 & 0.434 &     0.451 &2 & 
\tablenotemark{(g)}\\
 11 & F & 1.64127  &    17.579 &    17.149 & 0.430 &     0.442 &1 & 
 \\
 12 & RRc? & 0.30425  &    16.806 &    16.916 &-0.110 &    -0.110 &1 & 
 \\
 13 & F & 2.73870  &    16.707 &    16.186 & 0.521 &     0.526 &1 & 
 \\
 14 & O & 1.72946  &    16.930 &    16.525 & 0.405 &     0.410 &1 & 
 \\
 15 & F & 4.18143  &    16.546 &    16.019 & 0.527 &     0.557 &1 & 
 \\
 16 & O & 1.29851  &    16.821 &    16.505 & 0.316 &     0.321 &1 & 
 \\
 17 & E & 3.07225  &    16.469 &    16.543 &-0.074 &    -0.075 &1 & 
 \\
 18 & O & 1.58453  &    16.709 &    16.316 & 0.393 &     0.402 &1 & 
 \\
 19 & O & 1.20883  &    16.844 &    16.504 & 0.340 &     0.349 &1 & 
 \\
 20 & E & 2.56280  &    15.322 &    15.500 &-0.178 &    -0.179 &1 & 
 \\
 21 & O? & 1.06209  &    17.185 &    16.870 & 0.315 &     0.324 &2 & 
 \\
 22 & F? & 3.93866  &    16.808 &    16.513 & 0.295 &     0.298 &2 & 
\tablenotemark{(h)}\\
 23 & F & 1.16736  &    18.462 &    18.024 & 0.438 &     0.453 &2 & 
 \\
 24 & O? & 0.86912  &    17.501 &    17.175 & 0.326 &     0.337 &2 & 
 \\
 25 & E & 1.47070  &    15.522 &    15.580 &-0.058 &    -0.058 &2 & 
 \\
 26 & O & 2.41914  &    15.950 &    15.651 & 0.299 &     0.303 &1 & 
 \\
 27 & O? & 0.89960  &    17.908 &    17.438 & 0.470 &     0.476 &2 & 
 \\
 28 & O & 1.25427  &    17.041 &    16.688 & 0.353 &     0.362 &1 & 
 \\
 29 & RRab & 0.57382  &    19.860 &    19.498 & 0.362 &     0.391 &1 & 
 \\
 30 & F & 3.57270  &    16.925 &    16.331 & 0.594 &     0.611 &1 & 
 \\
 31 & F & 1.87530  &    17.838 &    17.292 & 0.546 &     0.549 &1 & 
 \\
 32 & O & 1.13510  &    17.227 &    16.884 & 0.343 &     0.349 &2 & 
 \\
 33 & F & 2.73273  &    17.348 &    16.801 & 0.547 &     0.551 &1 & 
 \\
 34 & F & 1.51413  &    17.892 &    17.435 & 0.457 &     0.469 &2 & 
 \\
 35 & E & 0.95080  &    17.913 &    18.009 &-0.096 &    -0.096 &2 & 
\tablenotemark{(i)}\\
 36 & F & 4.87578  &    16.304 &    15.729 & 0.575 &     0.602 &1 & 
 \\
 37 & E & 1.78000  &    18.213 &    16.739 & 1.474 &     1.472 &2 & 
 \\
 38 & F & 1.30524  &    18.691 &    18.243 & 0.448 &     0.452 &2 & 
 \\
 39 & F & 1.87157  &    17.463 &    16.999 & 0.464 &     0.479 &1 & 
 \\
 40 & F & 1.87780  &    17.316 &    16.846 & 0.470 &     0.473 &1 & 
 \\
 41 & O & 1.05237  &    17.407 &    17.067 & 0.340 &     0.355 &2 & 
 \\
 42 & O & 0.97429  &    16.613 &    16.461 & 0.152 &     0.154 &2 & 
\tablenotemark{(j)}\\
 43 & O & 1.92178  &    16.508 &    16.093 & 0.415 &     0.417 &1 & 
 \\
 44 & RRab & 0.62042  &    19.705 &    19.347 & 0.358 &     0.382 &1 & 
 \\
 45 & O & 1.13099  &    17.064 &    16.770 & 0.294 &     0.306 &1 & 
 \\
 46 & O & 2.44532  &    16.324 &    15.908 & 0.416 &     0.421 &1 & 
 \\
 47 & O & 1.18244  &    17.165 &    16.875 & 0.290 &     0.303 &1 & 
 \\
 48 & E & 1.25565  &    17.018 &    17.119 &-0.101 &    -0.102 &2 & 
 \\
 49 & RRab & 0.53193  &    19.791 &    19.183 & 0.608 &     0.632 &1 & 
 \\
 50 & O & 2.31180  &    16.410 & \nodata &  \nodata & \nodata &1 & 
 \\
 51 & O & 1.76335  &    16.537 &    16.133 & 0.404 &     0.415 &1 & 
 \\
 52 & O & 1.02323  &    16.948 &    16.764 & 0.184 &     0.182 &2 & 
\tablenotemark{(k)}\\
 53 & F & 1.32225  &    17.962 &    17.416 & 0.546 &     0.554 &2 & 
 \\
 54 & O & 1.38819  &    16.690 &    16.390 & 0.300 &     0.304 &1 & 
 \\
 55 & O & 1.05222  &    17.200 &    16.884 & 0.316 &     0.326 &2 & 
 \\
 56 & O & 1.71675  &    16.630 &    16.281 & 0.349 &     0.354 &1 & 
 \\
 57 & O & 2.45760  &    16.208 &    15.809 & 0.399 &     0.403 &1 & 
 \\
 58 & O & 3.15021  &    15.944 &    15.557 & 0.387 &     0.389 &1 & 
 \\
 59 & O & 2.31693  &    16.358 &    15.959 & 0.399 &     0.406 &1 & 
 \\
 60 & F & 1.55752  &    17.783 &    17.280 & 0.503 &     0.522 &1 & 
 \\
 61 & E & 1.07459  &    17.522 &    17.584 &-0.062 &    -0.064 &2 & 
 \\
 62 & F & 3.37341  &    17.028 &    16.409 & 0.619 &     0.623 &1 & 
 \\
 63 & F & 3.16083  &    16.882 &    16.403 & 0.479 &     0.484 &2 & 
 \\
 64 & O & 2.00913  &    16.482 &    16.072 & 0.410 &     0.412 &1 & 
 \\
 65 & F & 4.59607  &    16.720 &    16.058 & 0.662 &     0.687 &2 & 
 \\
 66 & O & 1.52643  &    17.139 &    16.731 & 0.408 &     0.412 &1 & 
 \\
 67 & F & 2.64359  &    16.667 &    16.192 & 0.475 &     0.481 &1 & 
 \\
 68 & O & 2.42039  &    16.381 &    15.964 & 0.417 &     0.423 &2 & 
 \\
 69 & O & 0.95101  &    18.082 &    17.625 & 0.457 &     0.463 &2 & 
\tablenotemark{(l)}\\
 70 & F & 1.36882  &    17.924 &    17.522 & 0.402 &     0.410 &1 & 
 \\
 71 & F & 5.48753  &    16.518 &    15.909 & 0.609 &     0.612 &1 & 
 \\
 72 & O & 1.35233  &    17.116 &    16.756 & 0.360 &     0.367 &1 & 
 \\
 73 & E & 0.79765  & \nodata &     18.489 & \nodata & \nodata &2 & 
 \\
 74 & O & 2.24879  &    16.613 &    16.143 & 0.470 &     0.476 &1 & 
 \\
 75 & O & 3.07698  &    15.964 &    15.578 & 0.386 &     0.389 &2 & 
 \\
 76 & O & 0.83412  &    17.613 &    17.282 & 0.331 &     0.336 &1 & 
 \\
 77 & F & 3.16419  &    16.802 &    16.345 & 0.457 &     0.490 &2 & 
 \\
 78 & RRab & 0.56426  &    19.992 &    19.697 & 0.295 &     0.317 &2 & 
 \\
 79 & O & 2.40705  &    16.401 &    16.009 & 0.392 &     0.398 &2 & 
 \\
 80 & O & 1.17509  &    17.142 &    17.203 &-0.061 &    -0.053 &1(B) & 
 \\
 81 & RRab & 0.49061  & \nodata &     19.571 & \nodata & \nodata &1 & 
 \\
 82 & E & 1.52010  &    17.547 &    17.069 & 0.478 &     0.506 &2 & 
 \\
 83 & O & 2.11747  &    16.681 &    16.186 & 0.495 &     0.497 &1 & 
 \\
 84 & F & 6.06841  &    16.347 &    15.718 & 0.629 &     0.637 &2 & 
 \\
\enddata
\tablenotetext{(a)}{\protect{\citet{PGG66}} give P (days)=0.527340 and classify as possible RRab} 
\tablenotetext{(b)}{Probable blue companion \protect{\citep{A60}}} 
\tablenotetext{(c)}{\protect{\citet{PGG66}} classify as possible irregular, typing here indeterminate} 
\tablenotetext{(d)}{Much brighter here ($\langle B \rangle = 18.528$) than in  \protect{\citet{S92}} ($\langle B \rangle = 20.1$) } 
\tablenotetext{(e)}{Non-periodic?} 
\tablenotetext{(f)}{Possible blue companion?} 
\tablenotetext{(g)}{Noisy light curve makes typing uncertain} 
\tablenotetext{(h)}{Noise in curves mimicking periodicity?} 
\tablenotetext{(i)}{Blue companion?, Alternate period at 0.47541 days} 
\tablenotetext{(j)}{Blue companion?} 
\tablenotetext{(k)}{Blue companion?} 
\tablenotetext{(l)}{Extremely low amplitude in $V$, curve morphology in $B$ more indicative of fundamental mode} 
\end{deluxetable}
 
\clearpage

\begin{deluxetable}{lccc} 
\tablewidth{0pc}
\footnotesize
\tablecaption{Reddenings of RR Lyrae Stars \label{tbl7}}
\tablehead{
\colhead{Star} & \colhead{[Fe/H]} & \colhead{$\Delta S$} & \colhead{$E_{B-V}$}
          }
\startdata
HV12949 & -1.15 & 5.76 & 0.04 \\
HV2043 & -1.50 & 7.96 & 0.07 \\
HV2102 & -1.18 & 5.97 & 0.06 \\
\textrm{[SSB92]} 26 & -0.84 & 3.79 & 0.09 \\
\textrm{[SSB92]} 29 & -1.03 & 4.97 & 0.05 \\
\textrm{[SSB92]} 89 & -0.93 & 4.37 & 0.01 \\
\enddata
\end{deluxetable}

\clearpage

\begin{deluxetable}{ccccc}
\tablewidth{0pc}
\tablecaption{Frequency Distribution of Samples: (a) Present Sample, (b) \protect{\citet{S92}} \label{tbl8}}
\tablehead{
\multicolumn{1}{l}{Period Range} & \multicolumn{2}{c}{N(fundamental)} & 
\multicolumn{2}{c}{N(first overtone)} \\
($\log_{10}$ days) & (a) & (b) & (a) & (b) 
}
\startdata
-0.3 -0.2 & 0 & 0 & 1 & 0 \\
-0.2 -0.1 & 0 & 0 & 4 & 6 \\
-0.1 0.0 & 1 & 1 & 15 & 8 \\
0.0 0.1 & 7 & 5 & 17 & 6 \\
0.1 0.2 & 20 & 15 & 11 & 6 \\
0.2 0.3 & 21 & 18 & 12 & 8 \\
0.3 0.4 & 15 & 9 & 16 & 4 \\
0.4 0.5 & 16 & 13 & 3 & 0 \\
0.5 0.6 & 15 & 8 & 1 & 1 \\
0.6 0.7 & 9 & 5 & 0 & 0 \\
0.7 0.8 & 9 & 6 & 0 & 1 \\
0.8 0.9 & 4 & 3 & 0 & 0 \\
0.9 1.0 & 1 & 2 & 0 & 0 \\
1.0 1.1 & 2 & 4 & 0 & 0 \\
1.1 1.2 & 1 & 0 & 0 & 0 \\
1.2 1.3 & 0 & 1 & 0 & 0 \\
\enddata
\end{deluxetable}

\clearpage

\begin{table}
\caption{Coefficients of $P-L$ relations. \label{tbl9}}
\begin{tabular}{cllll}
\tableline\tableline
\multicolumn{5}{l}{\textbf{Fundamental Mode}} \\ 
\multicolumn{2}{l}{76 Stars in B, 77 Stars in V} &
\multicolumn{1}{c}{$\alpha$} &
\multicolumn{1}{c}{$\beta$} \\ 
$\langle{\!B\!}\rangle_o$ & least squares & -2.50($\pm$0.12) & 17.76($\pm$0.05) & 
$\sigma=0.23$ \\ 
 & least abs. deviation & -2.58 & 17.78 & $\sigma=0.19$ \\ 
$\langle{\!V\!}\rangle_o$ & least squares & -2.79($\pm$0.11) & 17.47($\pm$0.05) & 
$\sigma=0.20$ \\ 
 & least abs. deviation & -2.79 & 17.45 & $\sigma=0.16$ \\ 
\multicolumn{5}{l}{\textbf{First Overtone Mode (1-OT)}} \\
\multicolumn{5}{l}{62 Stars in B, 60 Stars in V} \\ 
$\langle{\!B\!}\rangle_o$ & least squares & -2.87($\pm$0.12) & 17.10($\pm$0.03) & 
$\sigma=0.16$ \\ 
 & least abs. deviation & -2.87 & 17.10 & $\sigma=0.12$ \\ 
$\langle{\!V\!}\rangle_o$ & least squares & -3.03($\pm$0.10) & 16.83($\pm$0.03) & 
$\sigma=0.14$ \\ 
 & least abs. deviation & -3.02 & 16.83 & $\sigma=0.11$ \\ \tableline
\end{tabular}
\end{table}

\clearpage

\begin{table}
\caption{$W_{BFM}$ vs.\ $\log P$ \label{tbl10}}
\begin{tabular}{cllll}
\tableline\tableline
\multicolumn{5}{l}{\textbf{Fundamental Mode}} \\
\multicolumn{2}{l}{77 Stars in $V$} &
\multicolumn{1}{c}{$\alpha$} &
\multicolumn{1}{c}{$\beta$} \\  
$W_{BFM}$ & least squares & -3.63($\pm$0.12) & 16.42($\pm$0.05) & $\sigma=0.21$ \\ 

 & least abs. deviation & -3.55 & 16.39 & $\sigma=0.14$ \\ 
\multicolumn{5}{l}{\textbf{First Overtone Mode}} \\
\multicolumn{5}{l}{60 Stars in $V$} \\ 
$W_{BFM}$ & least squares & -3.41($\pm$0.12) & 15.88($\pm$0.03) & $\sigma=0.16$ \\ 

 & least abs. deviation & -3.56 & 15.91 & $\sigma=0.12$ \\ \tableline
\end{tabular}
\end{table}

\clearpage

\begin{table}
\caption{$P-L$ Relations for Different Ranges of Period \label{tbl11}} 
\begin{tabular}{clccl}
\tableline
\tableline
\multicolumn{2}{l}{\textbf{Fundamental Mode}} &
\multicolumn{1}{c}{$\alpha$} &
\multicolumn{1}{c}{$\beta$} \\
\multicolumn{5}{l}{Period $>$ 2 days (43 stars in $B$, 44 stars in $V$)} \\ 
$\langle{\!B\!}\rangle_{o}$ & least squares & -2.51($\pm$0.21) & 17.78($\pm$0.12) & 
$\sigma=0.23$ \\ 
 & least absolute deviation & -2.70 & 17.87 & $\sigma=0.18$ \\ 
$\langle{\!V\!}\rangle_{o}$ & least squares & -2.60($\pm$0.19) & 17.37($\pm$0.10) & 
$\sigma=0.19$ \\ 
 & least absolute deviation & -2.75 & 17.43 & $\sigma=0.15$ \\
\multicolumn{5}{l}{Period $<$ 2 days (33 stars in $B$, 33 stars in $V$)} \\ 

$\langle{\!B\!}\rangle_{o}$ & least squares & -3.47($\pm$0.44) & 17.92($\pm$0.09) & 
$\sigma=0.22$ \\ 
 & least absolute deviation & -3.86 & 17.98 & $\sigma=0.18$ \\ 
$\langle{\!V\!}\rangle_{o}$ & least squares & -3.86($\pm$0.37) & 17.67($\pm$0.08) & 
$\sigma=0.19$ \\ 
 & least absolute deviation & -3.56 & 17.58 & $\sigma=0.15$ \\ 
\multicolumn{5}{l}{\textbf{First Overtone Mode}} \\ 
\multicolumn{5}{l}{Period $>$ 1.4 days (32 stars in $B$, 31 stars in $V$)} \\ 

$\langle{\!B\!}\rangle_{o}$ & least squares & -3.04($\pm$0.35) & 17.18($\pm$0.11) & 
$\sigma=0.17$ \\ 
 & least absolute deviation & -3.27 & 17.27 & $\sigma=0.13$ \\ 
$\langle{\!V\!}\rangle_{o}$ & least squares & -3.12($\pm$0.31) & 16.88($\pm$0.10) & 
$\sigma=0.14$ \\ 
 & least absolute deviation & -3.02 & 16.84 & $\sigma=0.12$ \\ 
\multicolumn{5}{l}{Period $<$ 1.4 days (30 stars in $B$, 29 stars in $V$)} \\ 

$\langle{\!B\!}\rangle_{o}$ & least squares & -3.50($\pm$0.26) & 17.09($\pm$0.03) & 
$\sigma=0.14$ \\ 
 & least absolute deviation & -3.20 & 17.08 & $\sigma=0.11$ \\ 
$\langle{\!V\!}\rangle_{o}$ & least squares & -3.49($\pm$0.24) & 16.82($\pm$0.03) & 
$\sigma=0.14$ \\ 
 & least absolute deviation & -3.10 & 16.83 & $\sigma=0.10$ \\ \tableline 
\end{tabular}
\end{table}

\clearpage

\begin{table}
\caption{Coefficients of SMC $P-L$ Relations for Udalski et al.\ (1999a, as revised 2000) \label{tbl12}}
\begin{tabular}{lllllc}
\tableline
\tableline
\multicolumn{2}{l}{\textbf{Fundamental Mode}} 
& \multicolumn{1}{c}{$\alpha$} & \multicolumn{1}{c}{$\beta$} & & \multicolumn{1}{c}{N} \\ 
$\langle{\!V\!}\rangle_o$ & \textit{adopted} & -2.78 & 
17.64 & $\sigma=0.26$ & 464 \\ 
 & \textit{original} & -2.57($\pm$0.04) & 17.49($\pm$0.03) & 
$\sigma=0.25$ & 464 \\ 
$\langle{\!B\!}\rangle_o$ & \textit{original} & -2.21($\pm$0.05) & 17.72($\pm$0.04) &
$\sigma=0.32$ & 465 \\
\multicolumn{2}{l}{\textbf{First Overtone Mode}}
&  &  & &  \\
$\langle{\!V\!}\rangle_o$ & \textit{original} & -3.09($\pm$0.05) &
17.14($\pm$0.01) & $\sigma=0.26$ & 725 \\
$\langle{\!B\!}\rangle_o$ & \textit{original} & -2.93($\pm$0.06) &
17.44($\pm$0.02) & $\sigma=0.32$ & 729 \\
\tableline
\end{tabular}
\end{table}

\clearpage

\begin{table}
\caption{Coefficients of $P-L$ relations: Udalski et al.\ (1999b) Fields SC10 and SC11 \label{tbl13}}
\begin{tabular}{cllll}
\tableline \tableline
\multicolumn{2}{l}{\textbf{Fundamental Mode}} 
& \multicolumn{1}{c}{$\alpha$} & \multicolumn{1}{c}{$\beta$} & \\ 
\multicolumn{5}{l}{107 Stars in B, 107 Stars in V} \\ 
$\langle{\!B\!}\rangle_o$ & least squares & -2.67($\pm$0.10) & 17.85($\pm$0.06) & 
$\sigma=0.30$ \\ 
 & least abs. deviation & -2.76 & 17.91 & $\sigma=0.25$ \\ 
$\langle{\!V\!}\rangle_o$ & least squares & -2.91($\pm$0.08) & 17.56($\pm$0.05) & 
$\sigma=0.24$ \\ 
 & least abs. deviation & -2.91 & 17.58 & $\sigma=0.20$ \\ 
\multicolumn{5}{l}{\textbf{First Overtone Mode (1-OT)}} \\
\multicolumn{5}{l}{95 Stars in B, 95 Stars in V} \\ 
$\langle{\!B\!}\rangle_o$ & least squares & -3.04($\pm$0.18) & 17.27($\pm$0.05) & 
$\sigma=0.35$ \\ 
 & least abs. deviation & -3.07 & 17.28 & $\sigma=0.28$ \\ 
$\langle{\!V\!}\rangle_o$ & least squares & -3.20($\pm$0.15) & 17.01($\pm$0.04) & 
$\sigma=0.30$ \\  
 & least abs. deviation & -3.28 & 17.03 & $\sigma=0.24$ \\ \tableline
\end{tabular}
\end{table}

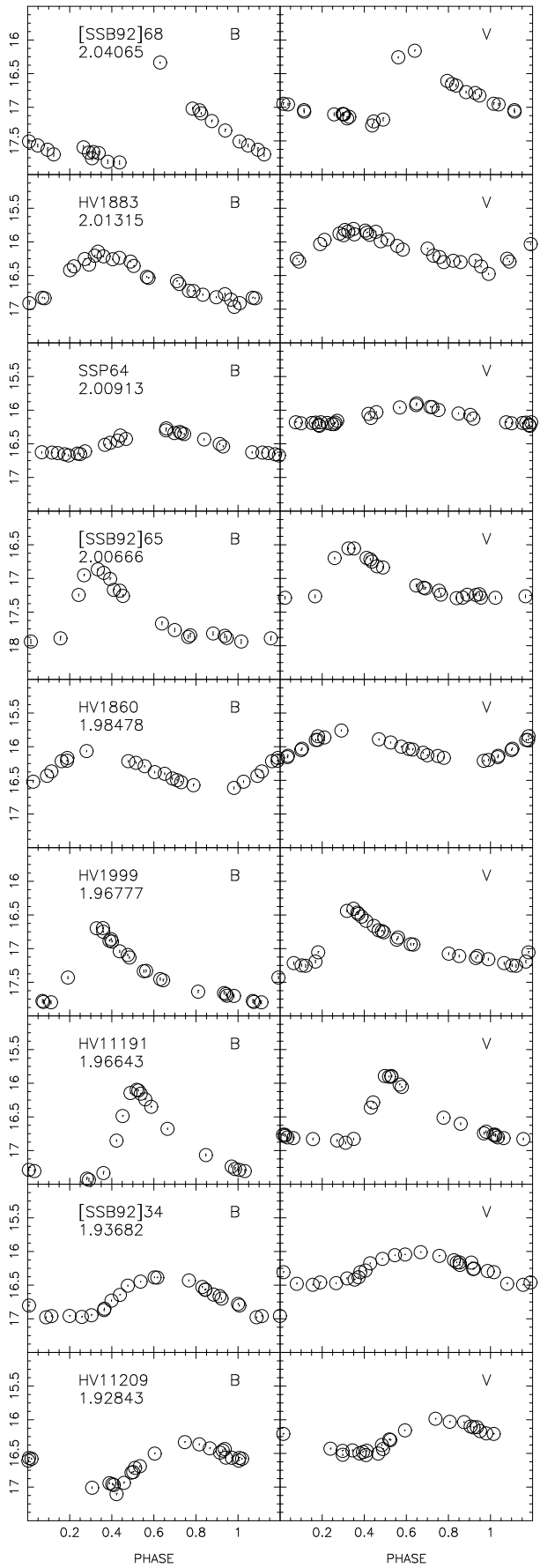
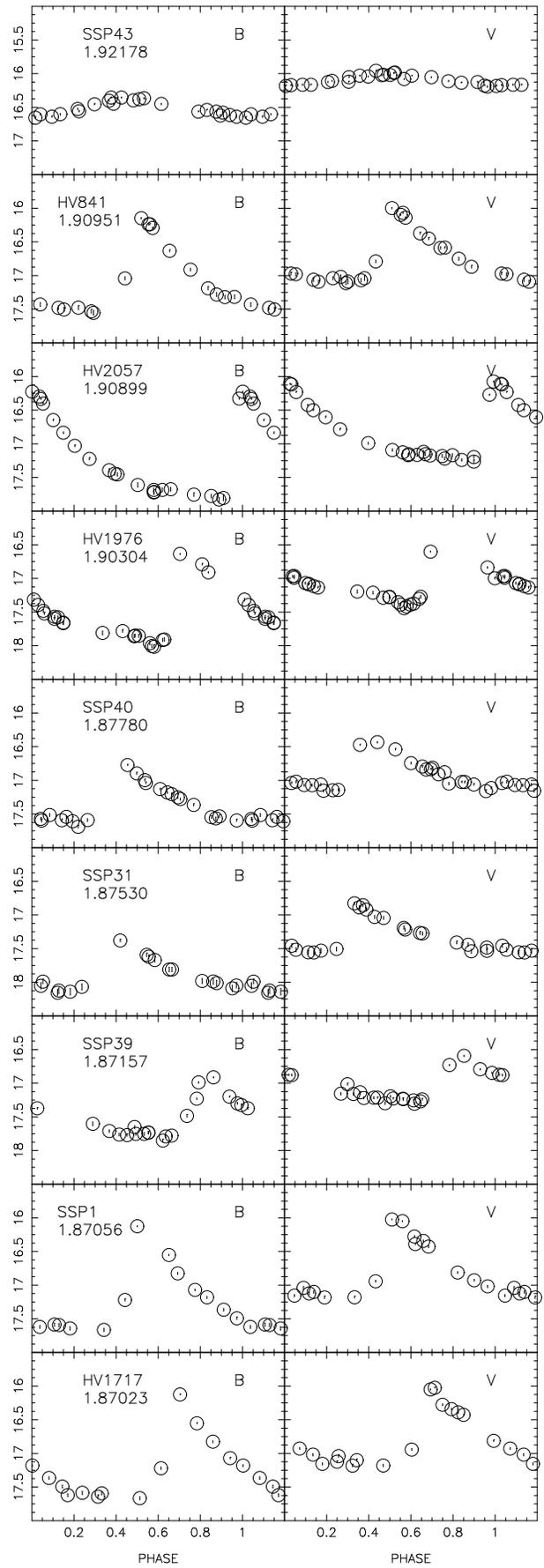
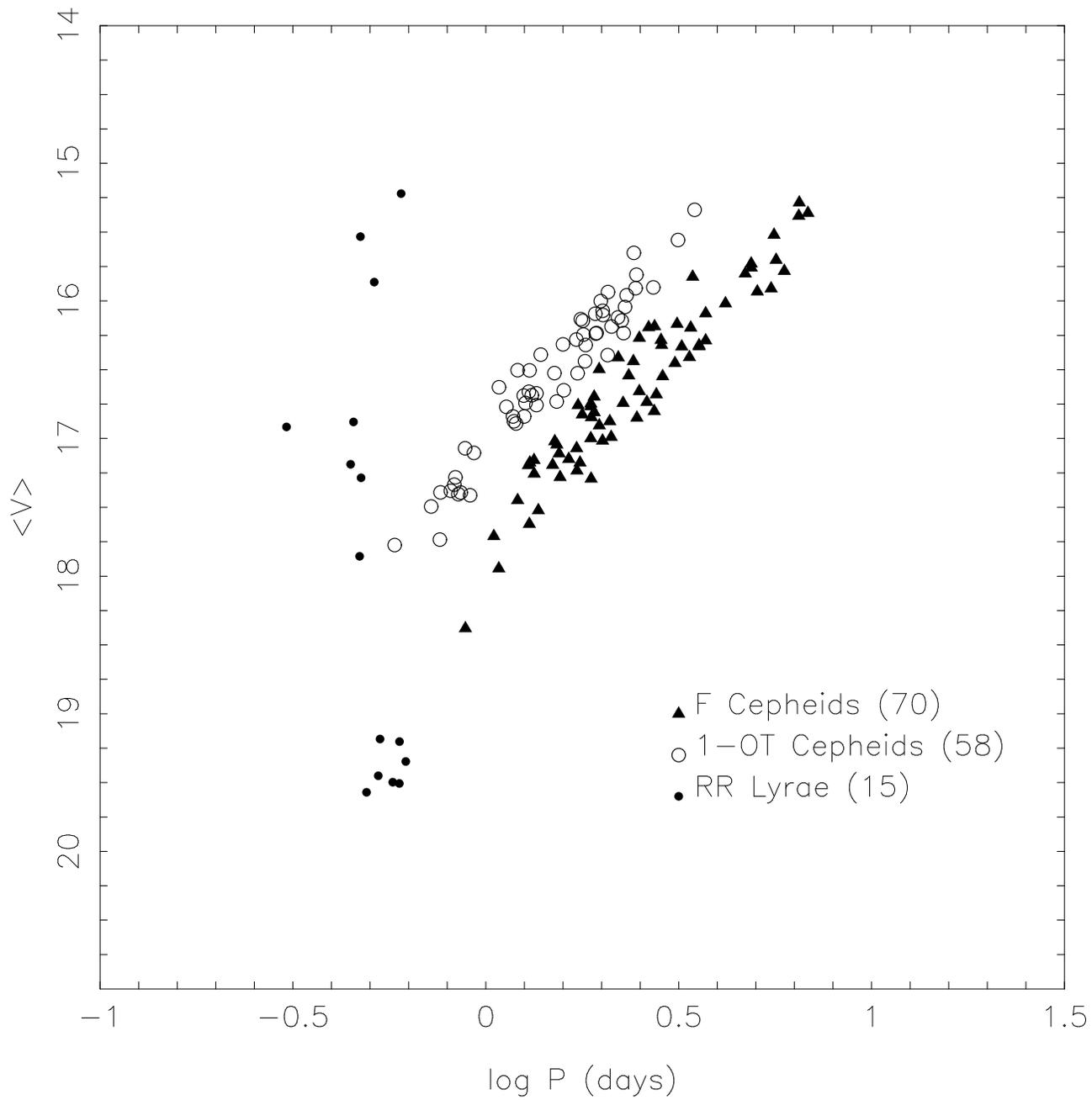
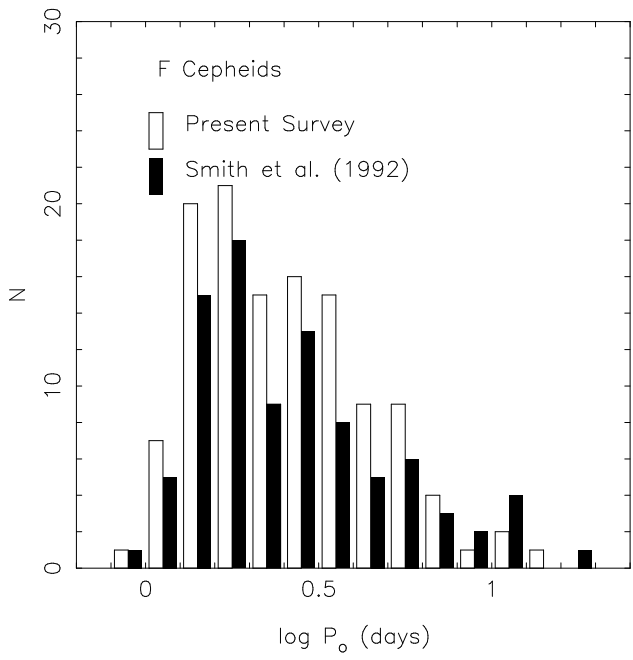
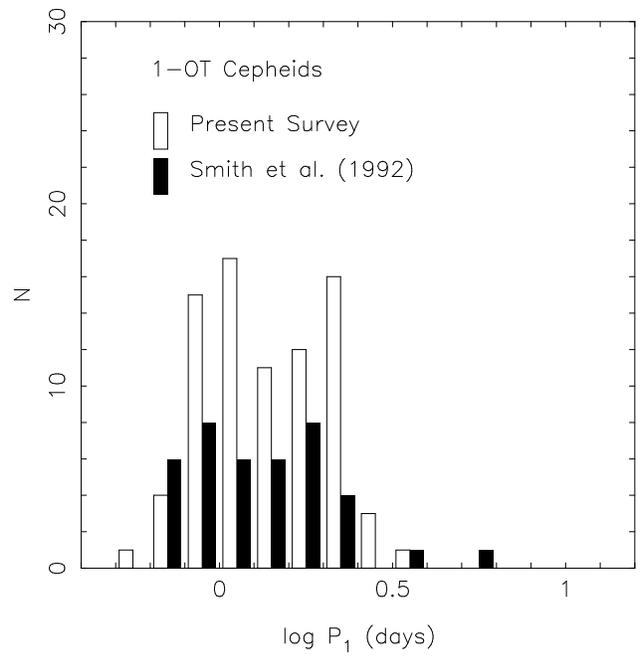
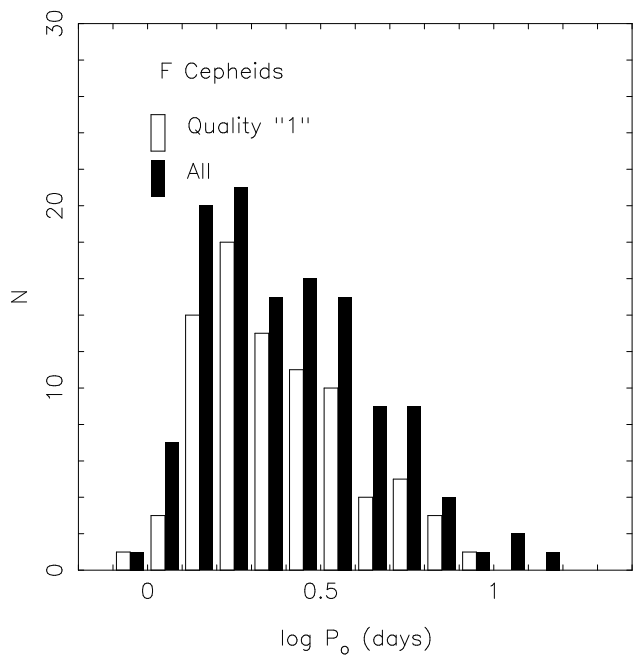
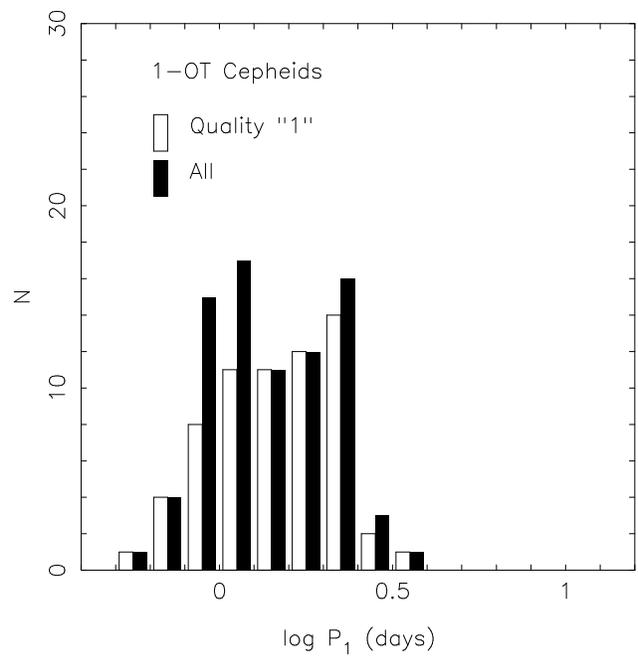
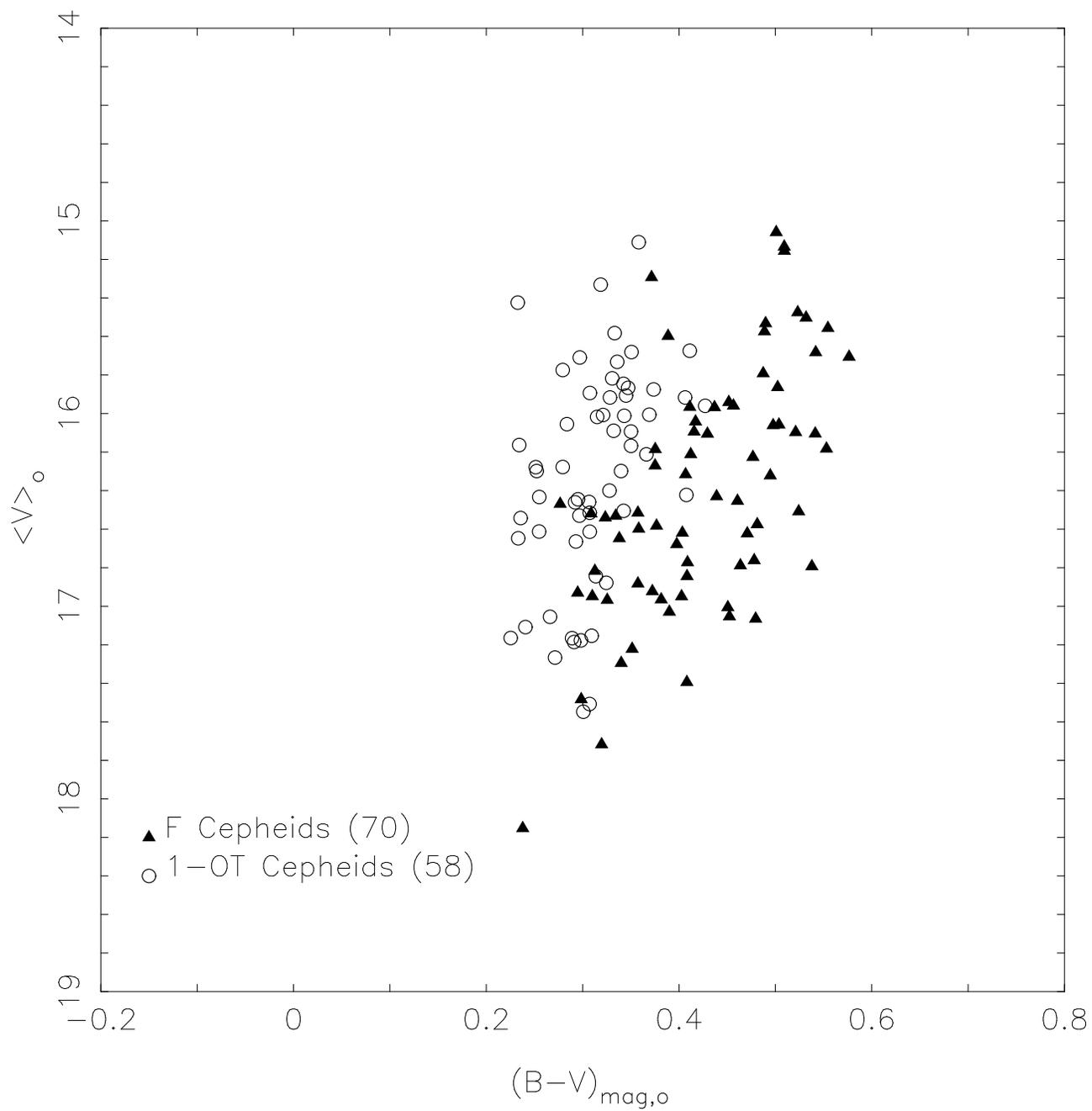
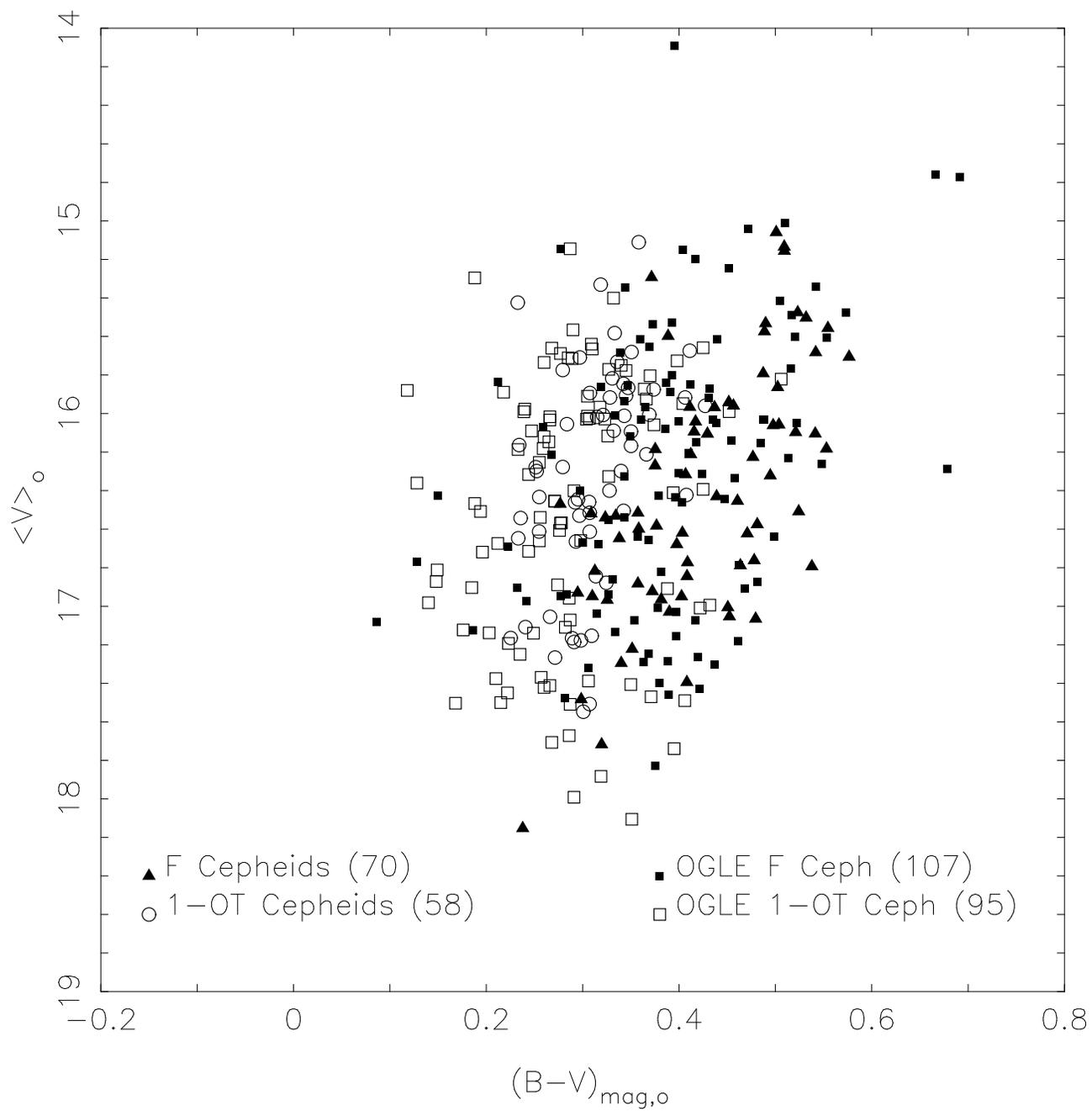
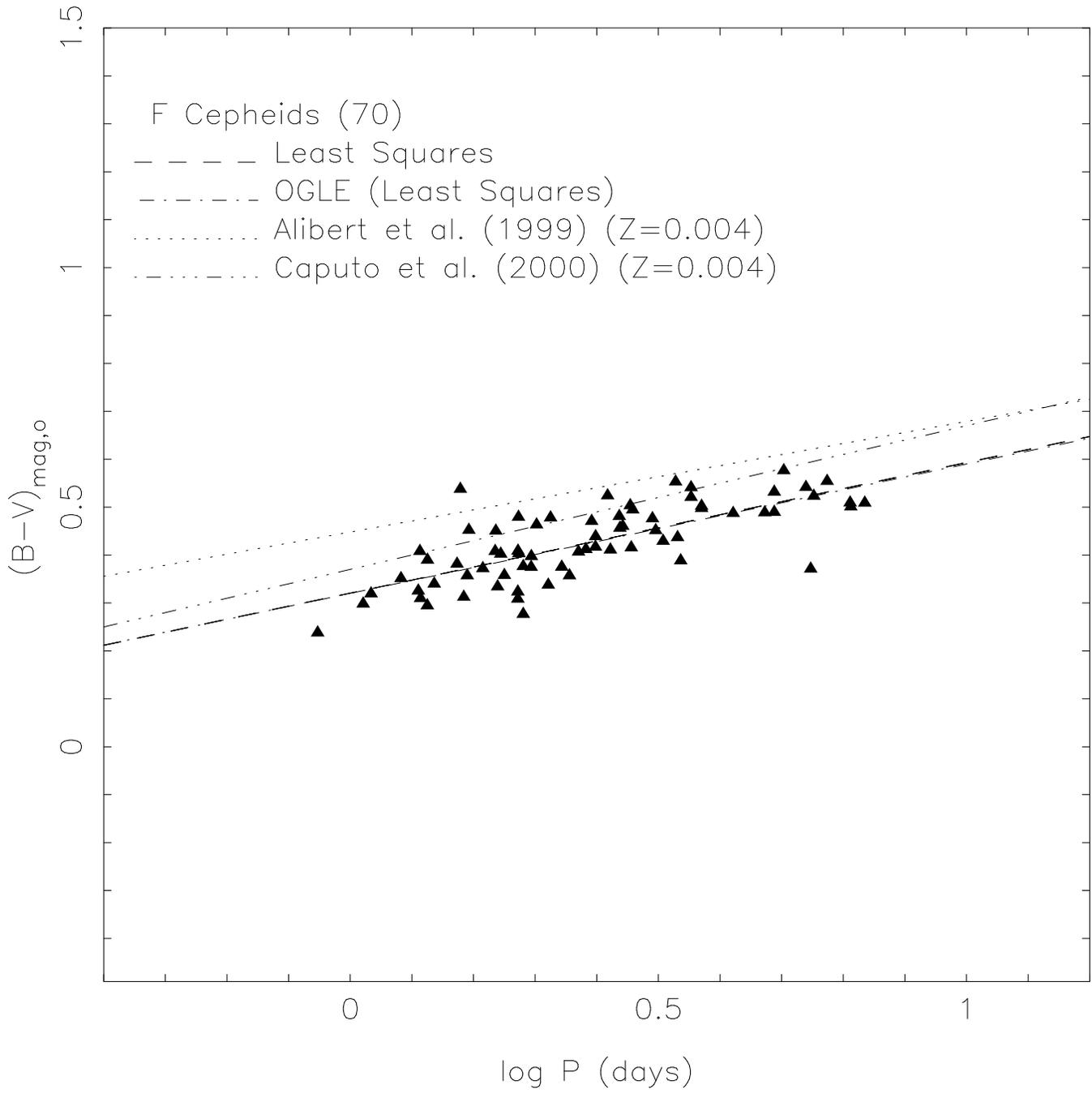
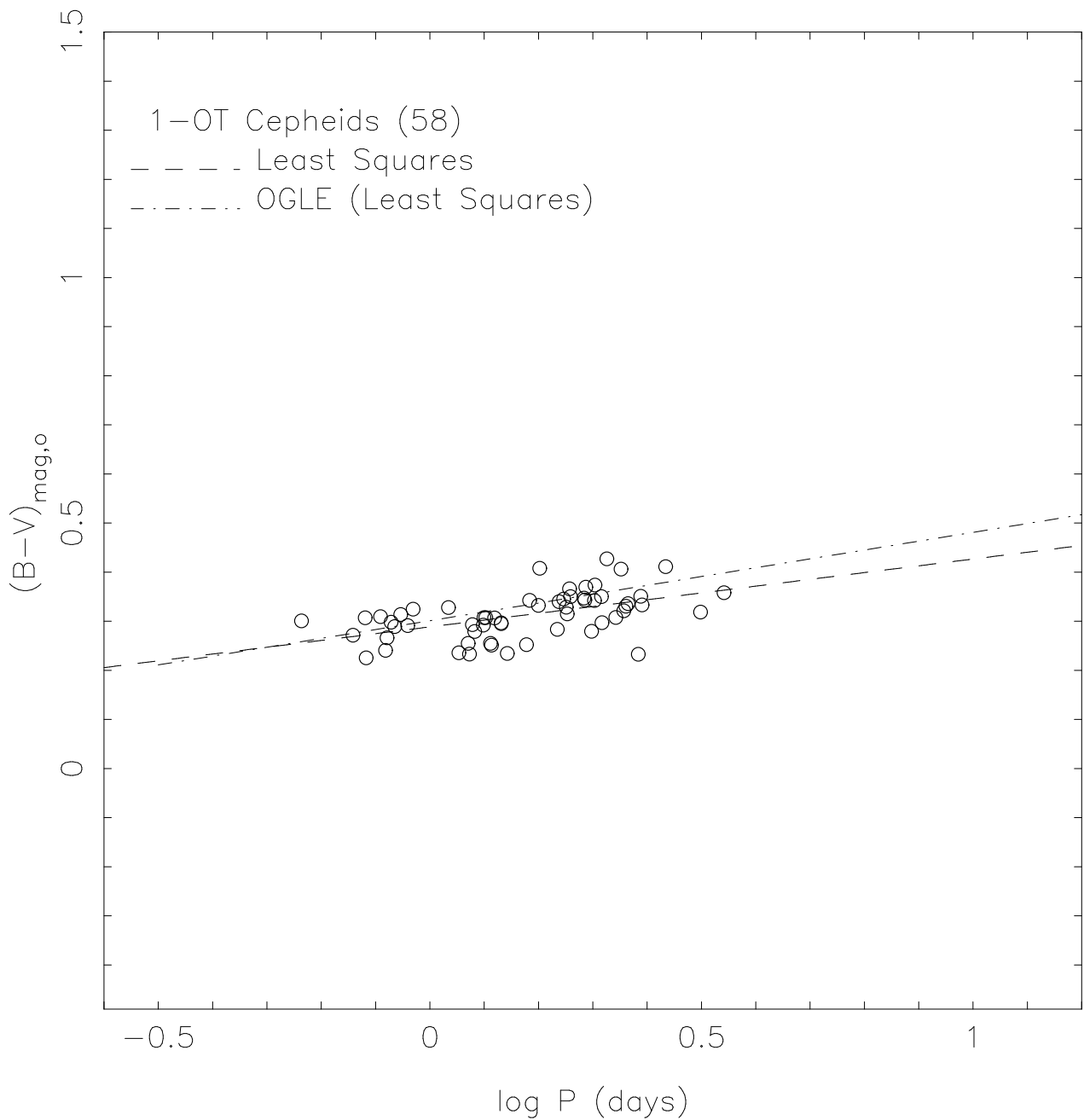
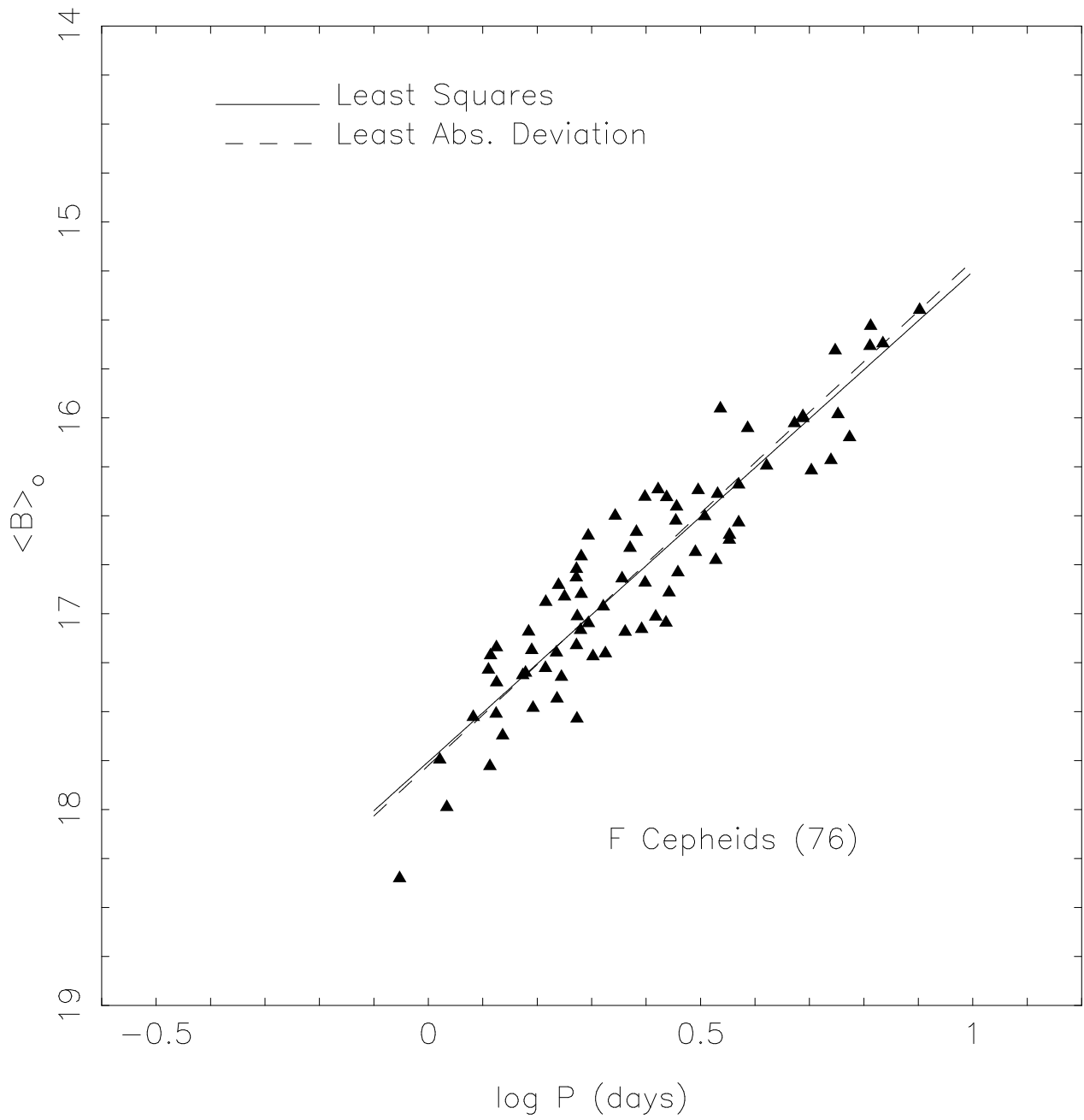
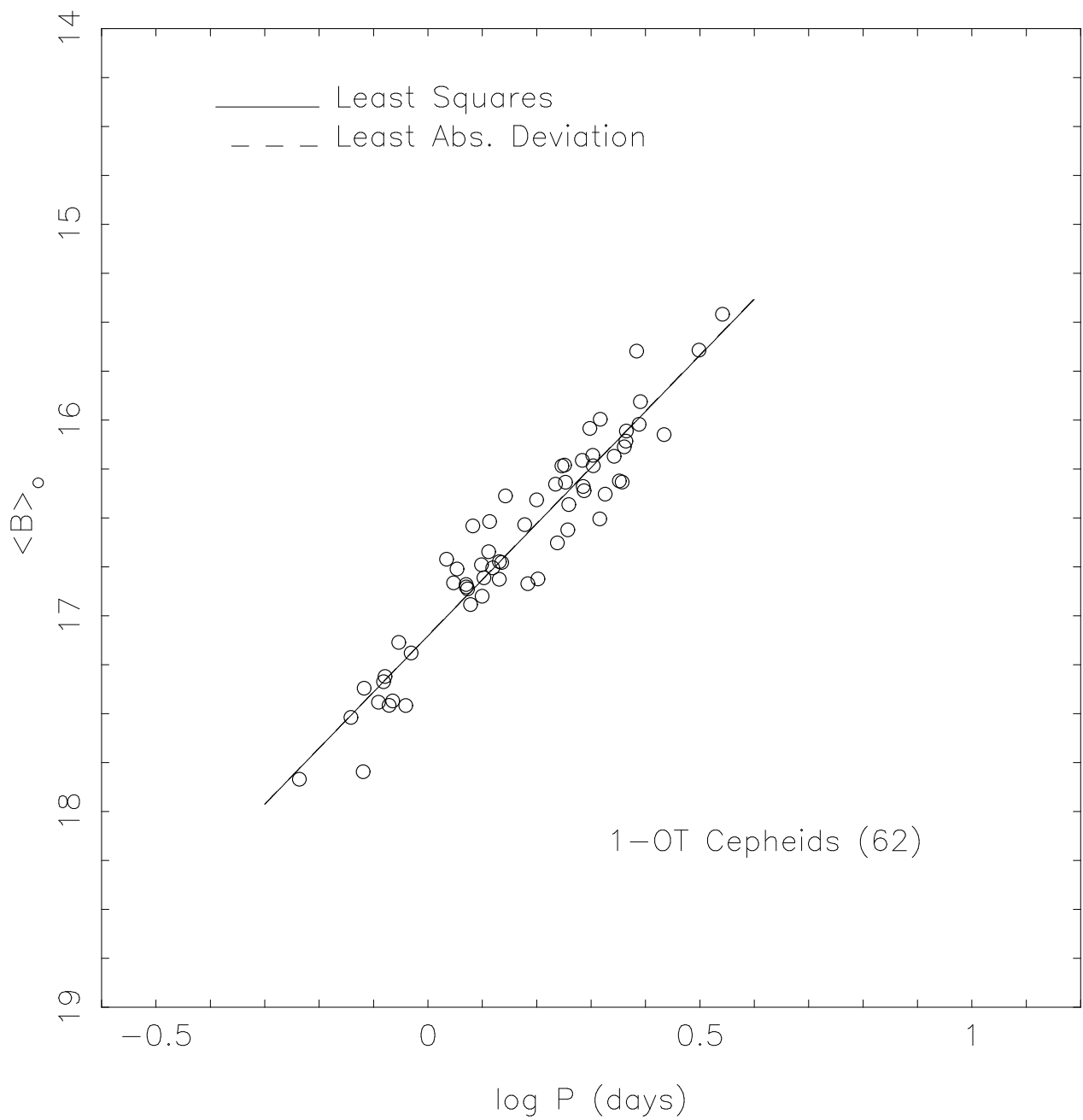
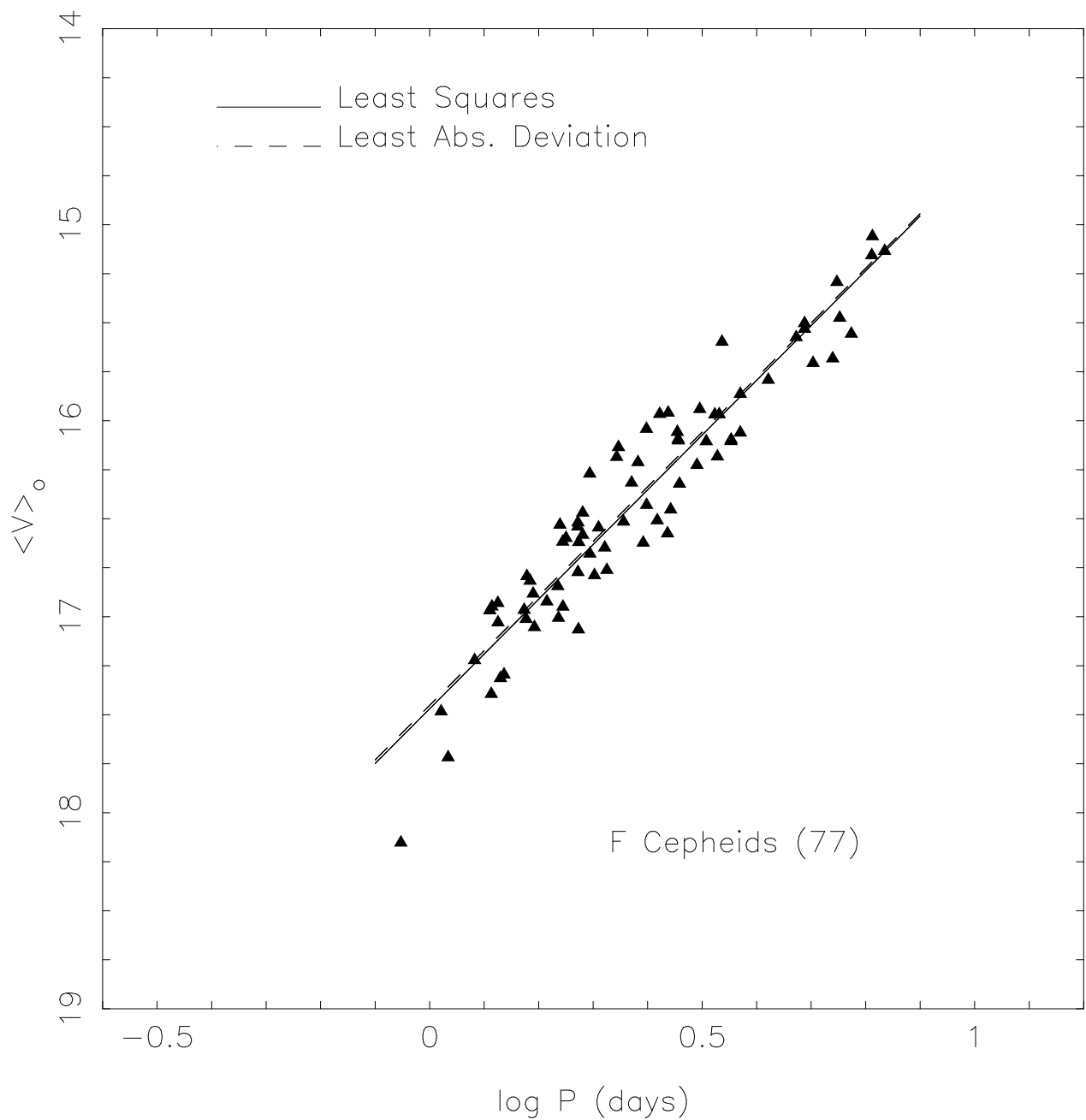
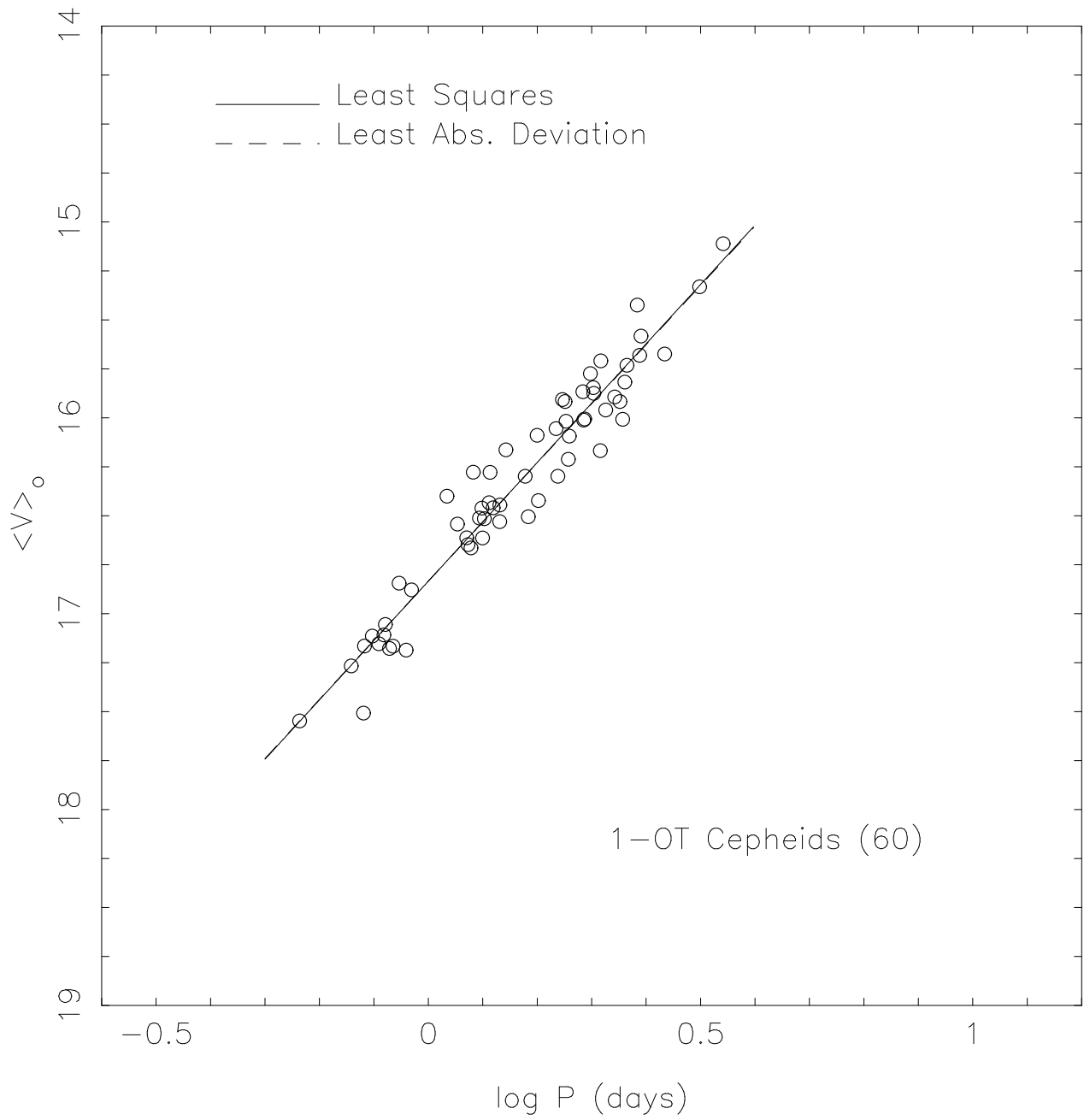
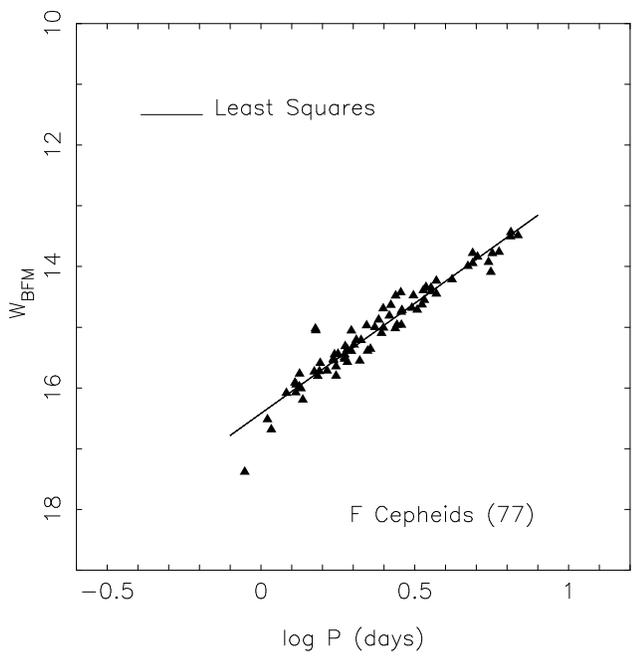
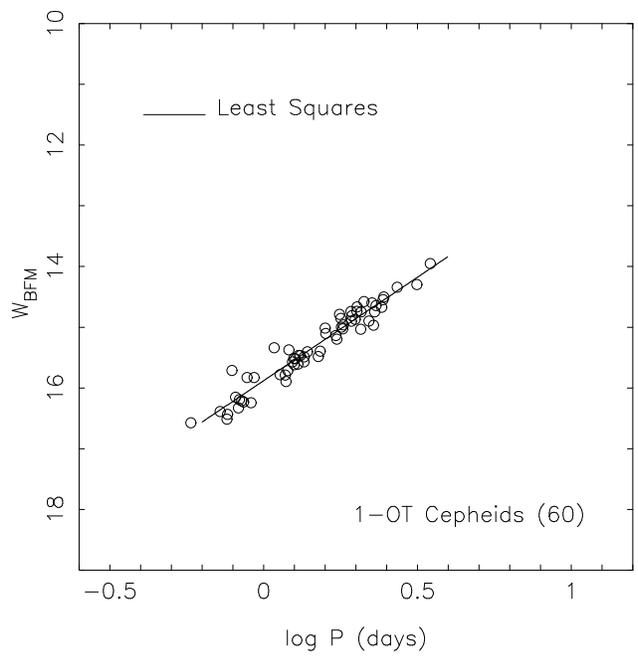
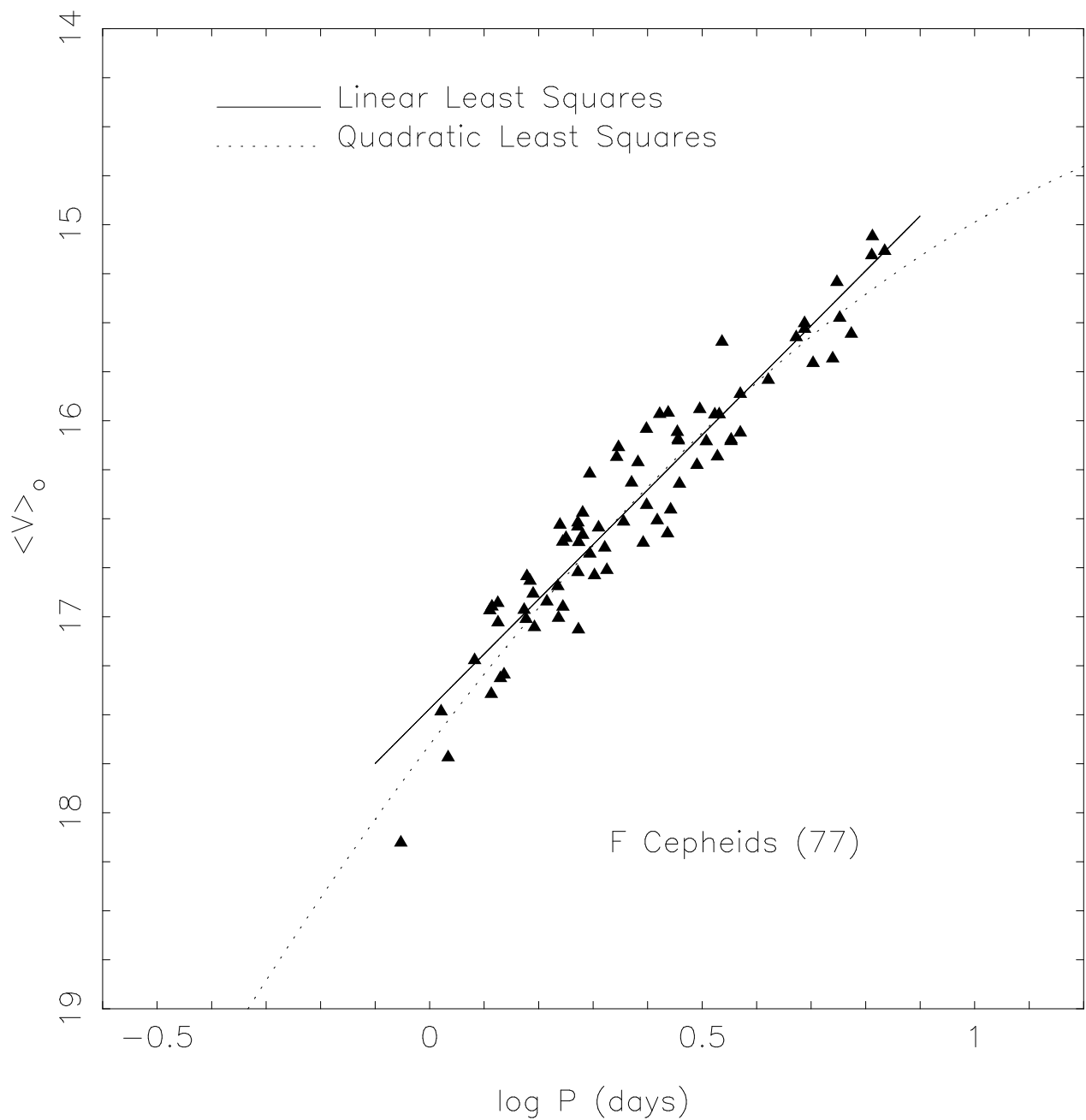
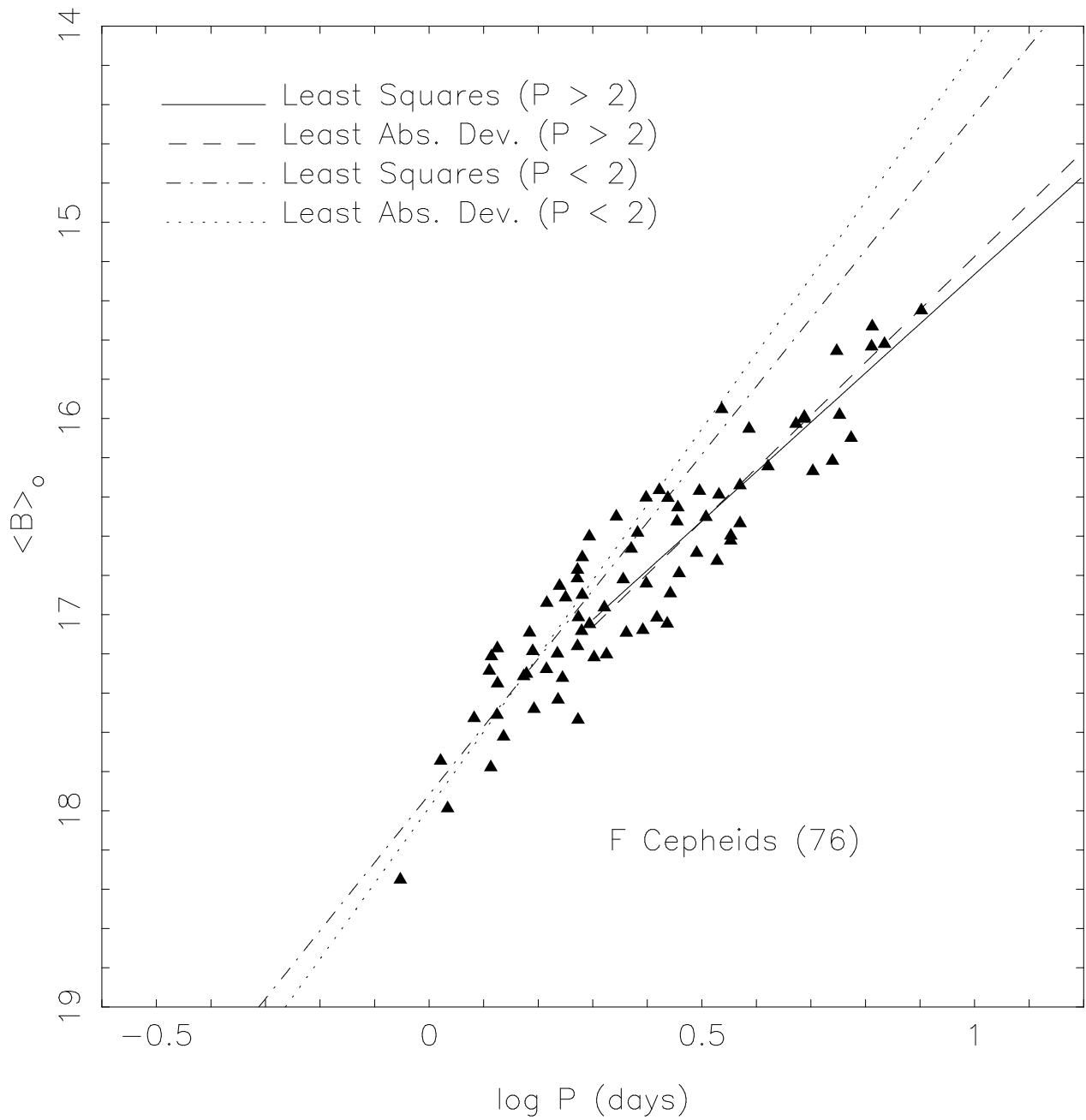
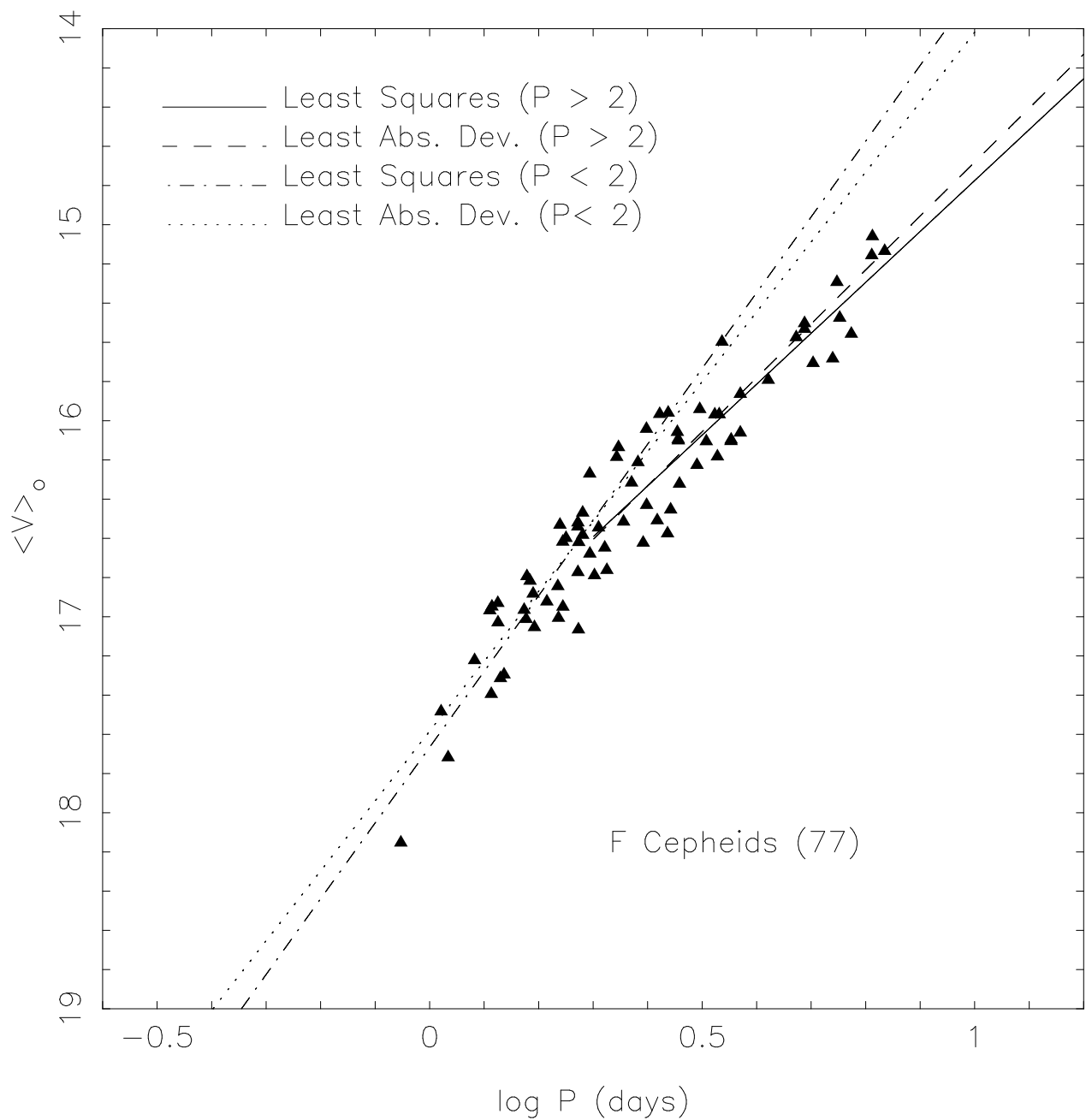
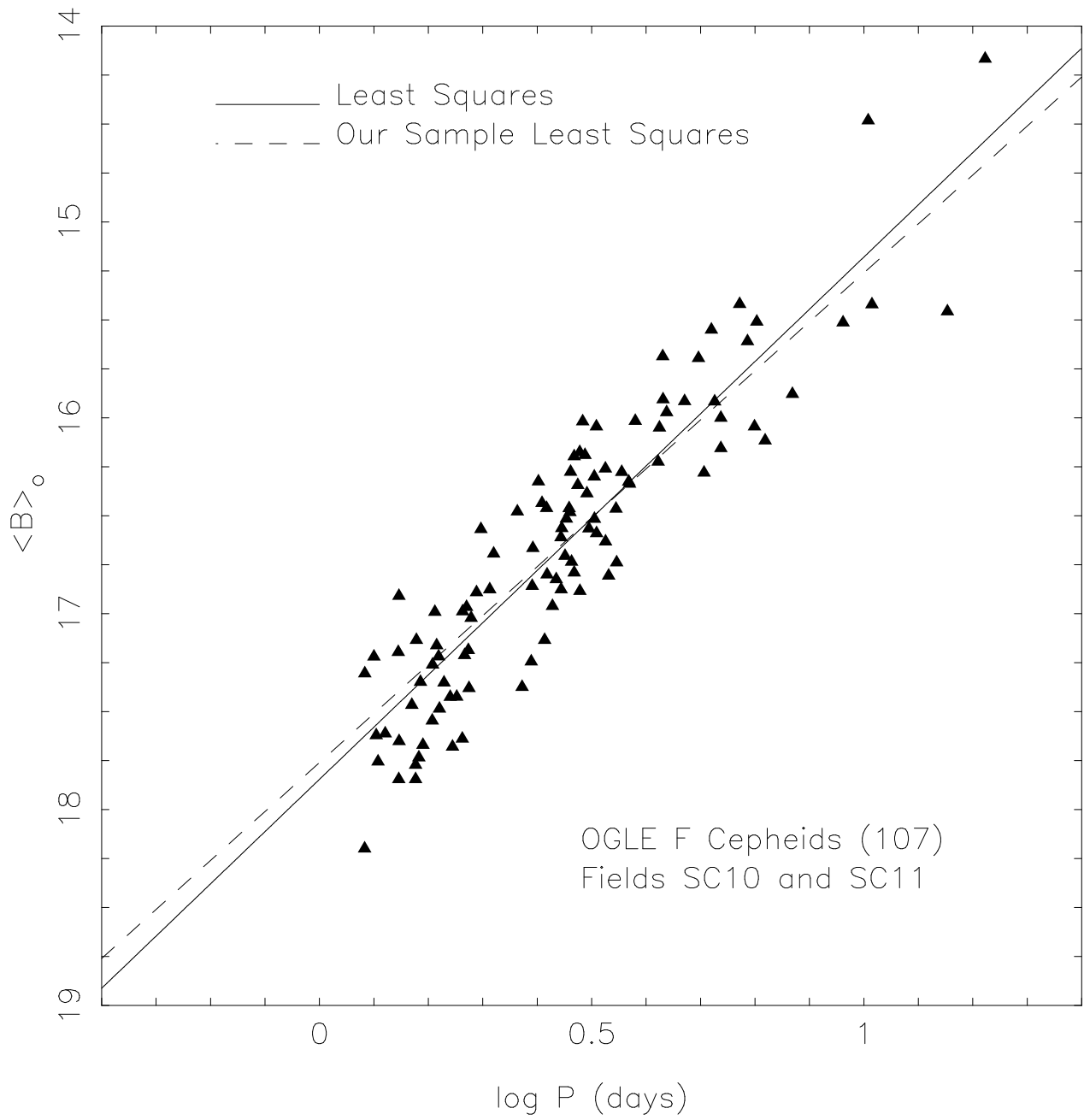
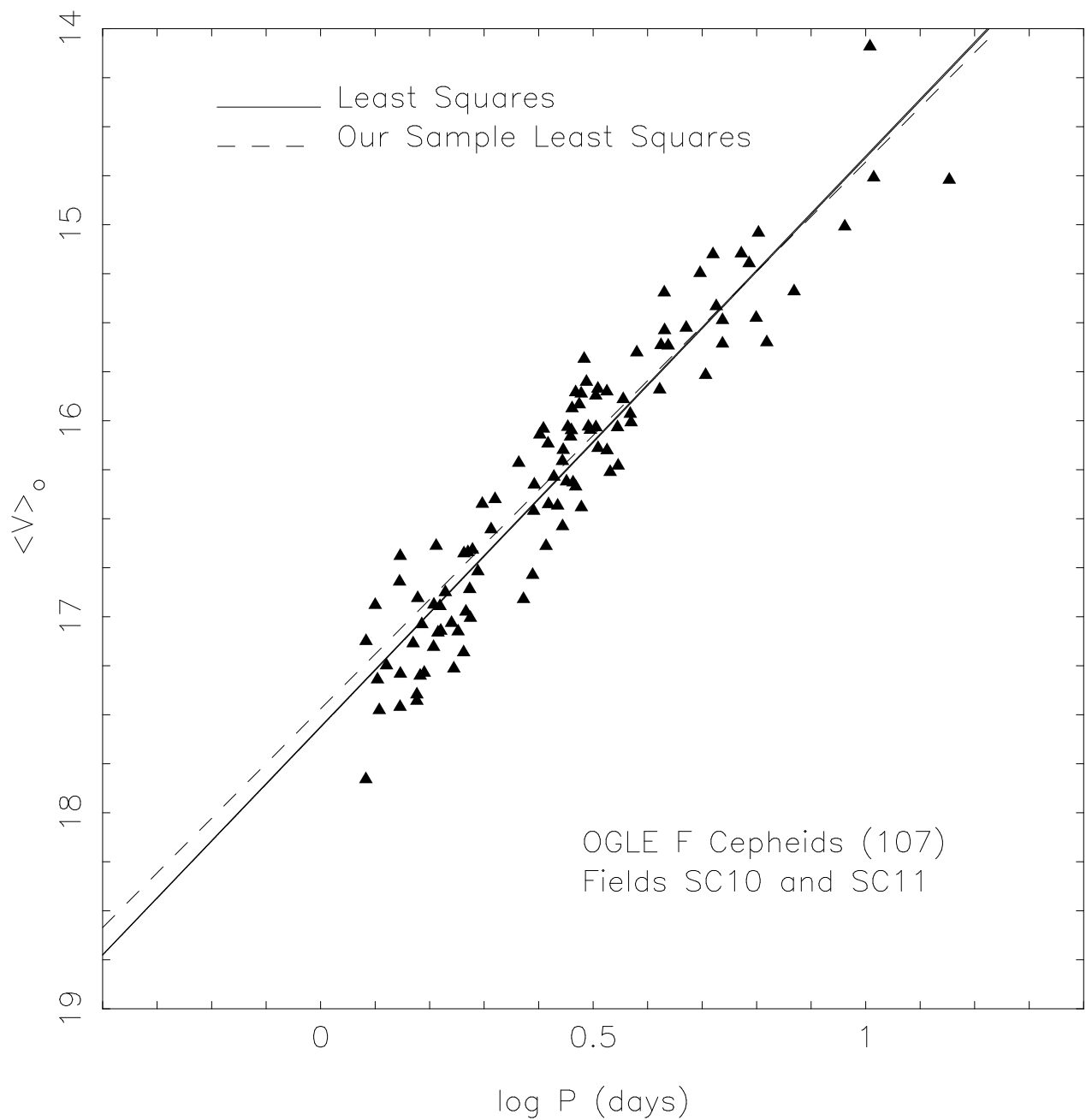
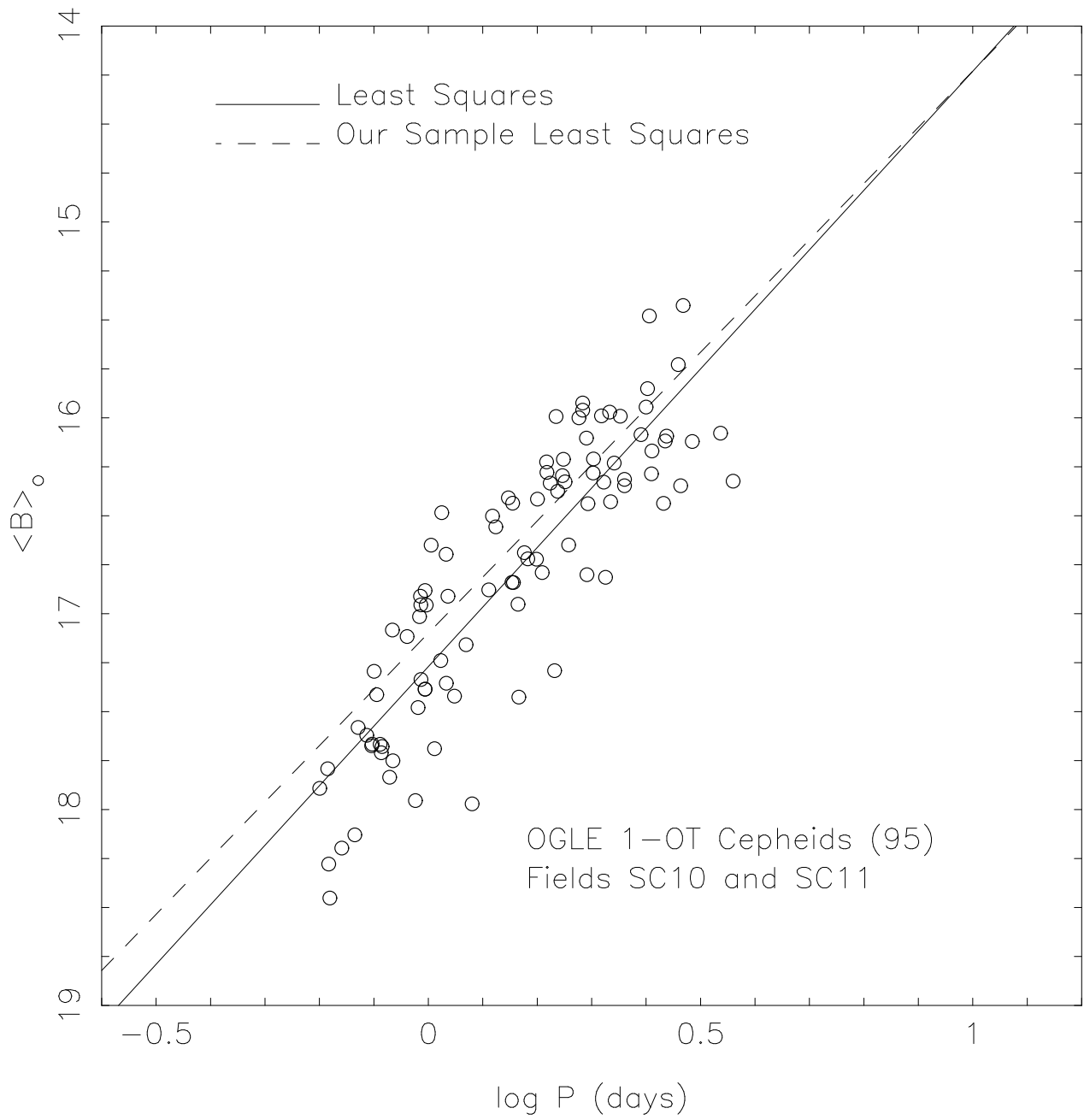
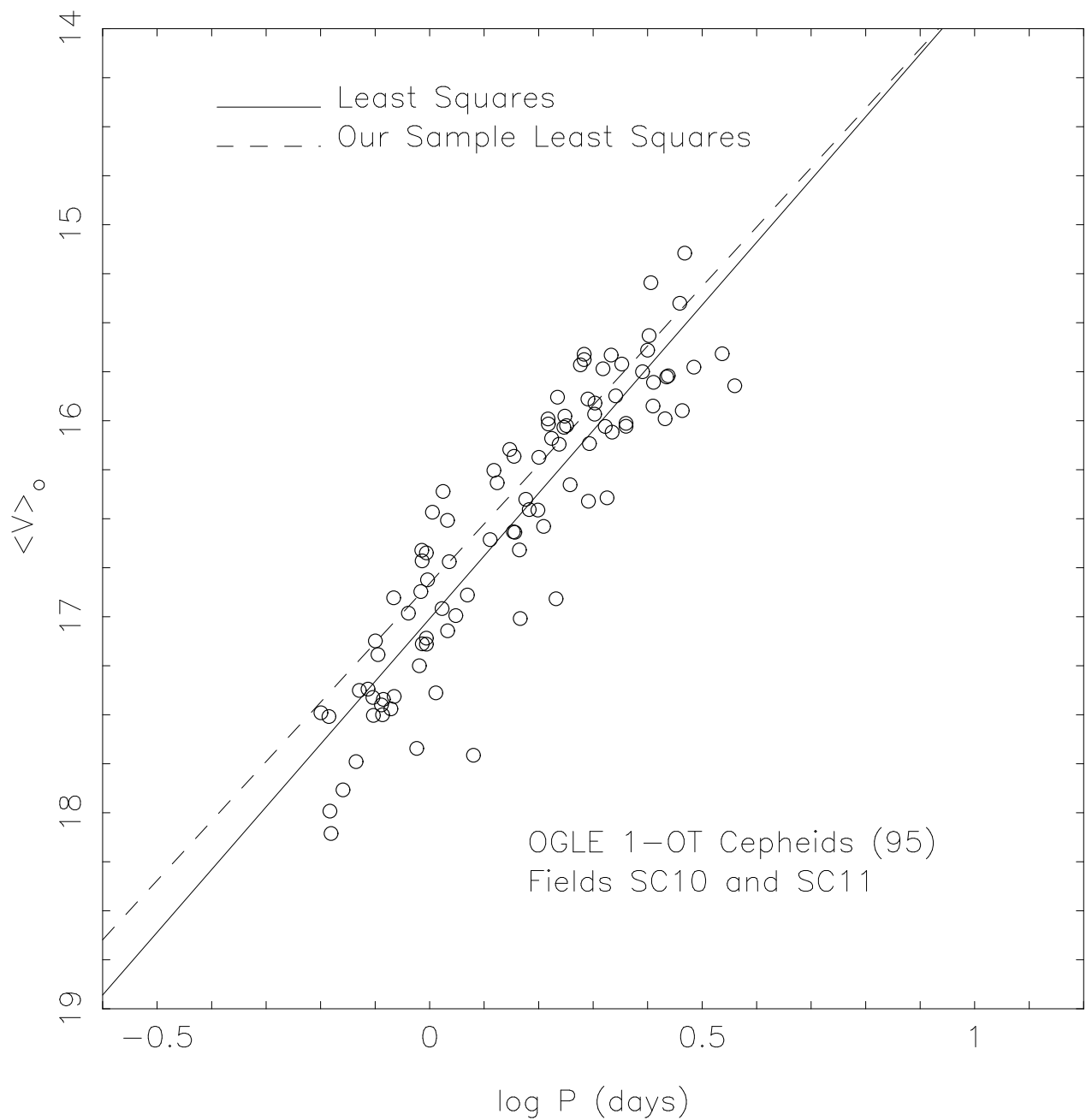
\end{document}